\documentclass[a4paper,11pt]{article}

\usepackage{jheppub} 

\usepackage[T1]{fontenc} 
\usepackage{graphicx}

\title{The collision of two-kinks revisited: the creation of kinks and lump-like defects as metastable states}

\author[a]{T. S. Mendon\c ca}

\author[a,1]{H. P. de Oliveira \note{Corresponding author.}}

\affiliation[a]{
Departamento de F\'{\i}sica Te\'orica - Instituto de F\'{\i}sica
A. D. Tavares, Universidade do Estado do Rio de Janeiro \\ 
R. S\~ao Francisco Xavier, 524. Rio de Janeiro, RJ, 20550-013, Brazil}

\emailAdd{tiagobrouwer@msn.com}
\emailAdd{hp.deoliveira@pq.cnpq.br}

\abstract{
We present a more detailed numerical investigation of the head-on collision of a two-kink/two-antikink system. We identified the escape of oscillon-like configurations as a pair of kinks of the standard $\phi^4$ model moving apart from each other. New pieces of evidence support that the lump-like defects can emerge from the two-kinks interaction to form metastable configurations. Moreover, these configurations signalize the windows of escape that have a fractal structure similar to the $n$-bounce sequence when the kinks of $\phi^4$ interact. As the last piece of the numerical experiment, we show that by perturbing conveniently a lump-like defect it is possible to recover another lump-like configuration as a metastable configuration. } 

\begin{document} 
\maketitle
\flushbottom

\section{Introduction}

Topological defects made up by a real scalar field in 1+1 dimensions are solutions of the Klein-Gordon

\begin{equation}
\frac{\partial^2 \phi}{\partial t^2} - \frac{\partial^2 \phi}{\partial x^2} + \frac{\partial V(\phi)}{\partial \phi} = 0, \label{eq1.1}
\end{equation}

\noindent with the requirement that the potential $V(\phi)$ possesses two or more degenerate minima. Therefore, depending on the potential, we can generate a large variety of topological defects or kink configurations for which the most celebrated is the $\phi^4$-model whose potential is $V(\phi) = (1-\phi^2)^2/2$. 

Kink solutions are spatially localized and non-perturbative configurations. This remarkable property allows us to use them to applications in several areas of physics such as condensed matter \cite{cond_matter}, fluid mechanics \cite{fluid}, particle physics \cite{particle_phys}, and cosmology \cite{cosmology}, to mention some. Of particular interest is the interaction of kinks represented by a collision of two kinks studied in several models, whose outcomes reveals remarkable nonlinear features. 

A crucial aspect of all kink solutions is their stability under small perturbations due to the conservation of their topological charge. There exists another class of solutions, the bell-shaped lumps-like defects identified as non-topological structures modeled by real scalar fields and governed by the one-dimensional Klein-Gordon equation  (\ref{eq1.1}), but these solutions are not stable under small perturbations \cite{bazeia,brihaye}. On the other hand, lump-like structures formed by complex scalar fields plays a relevant role several in several applications like in condensed matter \cite{lump_cond_matter}, to construct the q-balls \cite{q-balls,q-balls2,q-balls3} and to describe dark matter \cite{lump_dark}.

Recently, we investigated numerically the collision of a pair of  two-kinks \cite{tiago}, a new class of topological defects introduced by Bazeia et al. \cite{bazeia_2kinks}. The two-kinks arises from the scalar field model with the following potential:

\begin{equation}
V(\phi) = \frac{1}{2}\phi^2(\phi^{-\frac{1}{p}}-\phi^{\frac{1}{p}})^2, \label{eq1.2}
\end{equation}

\noindent where the parameter $p$ is related to the way the scalar field self-interact. For $p=1$ we recover the standard $\phi^4$ model while for $p=3,5,...$ we obtain the two-kinks configurations whose exact solutions are $\phi(x) \propto \tanh^p\left(x/p\right)$ \cite{bazeia_2kinks}. These solutions are stable under small perturbations due to the conservation of the topological charge as we have mentioned. If the parameter $p$ is even the solutions are the nontopological lump-like defects \cite{bazeia_2kinks} characterized by $\lim_{x \rightarrow \pm \infty}=\phi_{\mathrm{min}}$, where $\phi=\phi_{\mathrm{min}}$ denotes the single minimum of the potential. 

In this paper, we revisited the collision of a pair of two-kinks correcting some of the previous results and adding more relevant features about the critical configurations. We organized the work as follows. Section 2 shows the essential aspects of the two-kink collision with a description of the two main solutions and the role played by the internal modes of a two-kink. We remark that one of the new features is the formation and escape of kinks belonging to the $\phi^4$ model. In  Section 3 we discuss three connected issues: the critical configurations or solutions, the windows of escape and its the fractal structure. We investigate the possibility of generating critical configurations by disturbing conveniently a lump-like defect in Section 4. Finally, in Section 5 we present the final remarks.

\section{The collision of two-kinks revisited}

The outcome of the collision of a pair of two-kinks depends on the impact velocity once the initial position of both defects is fixed. We have considered the initial configuration \cite{aninos}

\begin{equation}
\phi_0(x,0) = -1 + \phi_K(x+x_0,0) + \phi_{\bar{K}}(x-x_0,0), \label{eq2.1}
\end{equation}

\noindent where $\phi_K(x+x_0,0)$ and $\phi_{\bar{K}}(x-x_0,0)$ are a two-kink and two-antikink located at $x=\mp$ respectively. They are initially moving towards each other with impact velocity $u$ provided that

\begin{equation}
\phi_{K,\bar{K}}(x \pm x_0,t) = \pm \tanh^p\left(\frac{x \pm x_0 - u t}{p \sqrt{1-u^2}}\right). \label{eq2.2}
\end{equation} 

\noindent We have fixed $x_0=15$ in all numerical experiments. Henceforth, we name kinks to the topological defect corresponding to $p=1$ or the $\phi^4$ model. We integrate the Klein-Gordon equation numerically with the spectral code introduced in Ref. \cite{tiago}. The code was optimized to run with higher truncation orders and therefore increasing the accuracy of the simulations. Briefly, the basic approximation of the scalar field is $\phi_a(t,x)=\sum_{k=0}^N\,a_k(t) \psi_k(x)$, where $a_k(t)$ represents the unknown modes, $\phi_k(x)$, $k-0,1..,N$ are the basis function and $N$ is the truncation order (cf. \cite{tiago}). In the present numerical experiments we set $N_{min}=500$ and $N_{max}=1000$ depending on the impact velocity.

\begin{figure}[htb]
\centering
\includegraphics[scale=0.5]{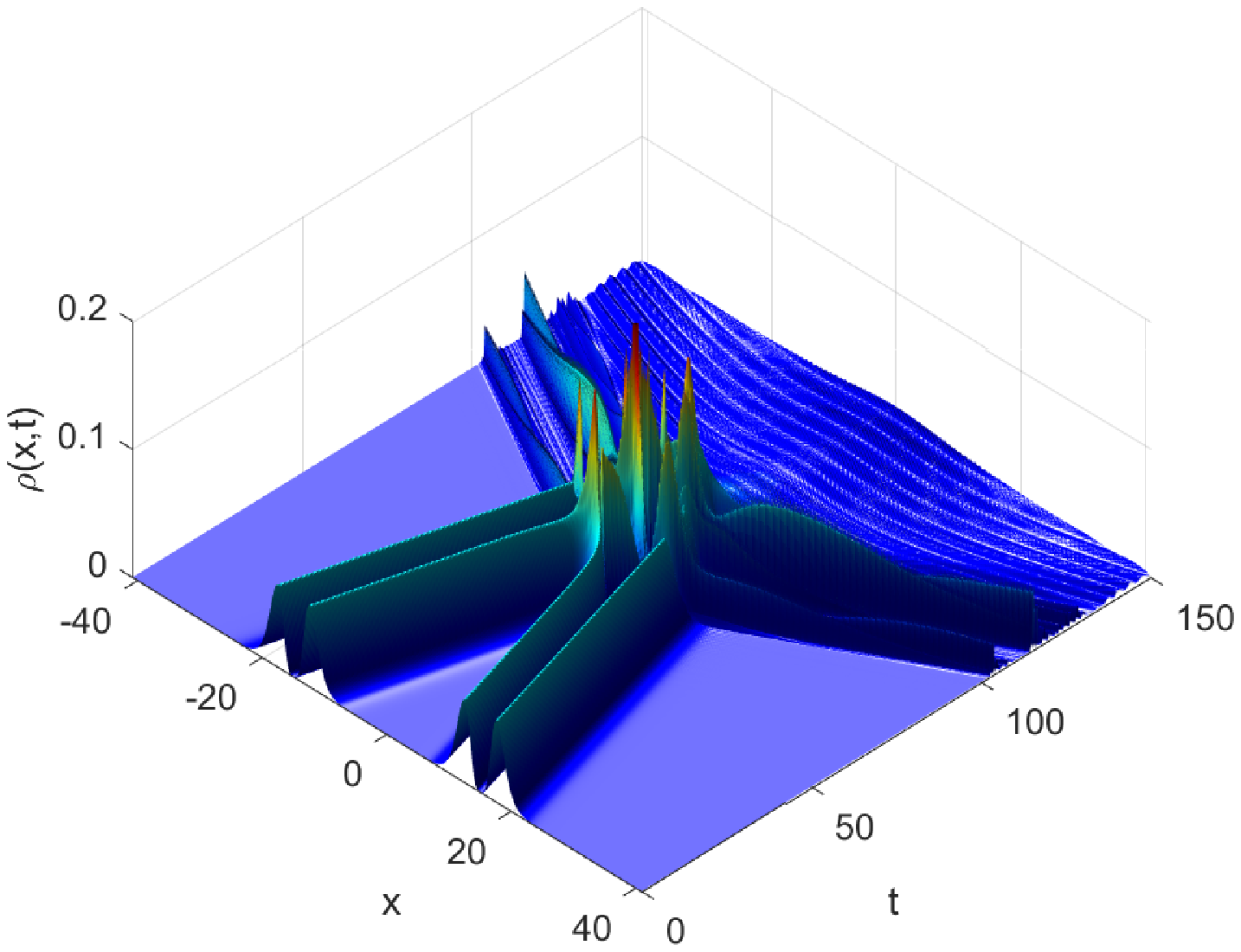}\includegraphics[scale=0.5]{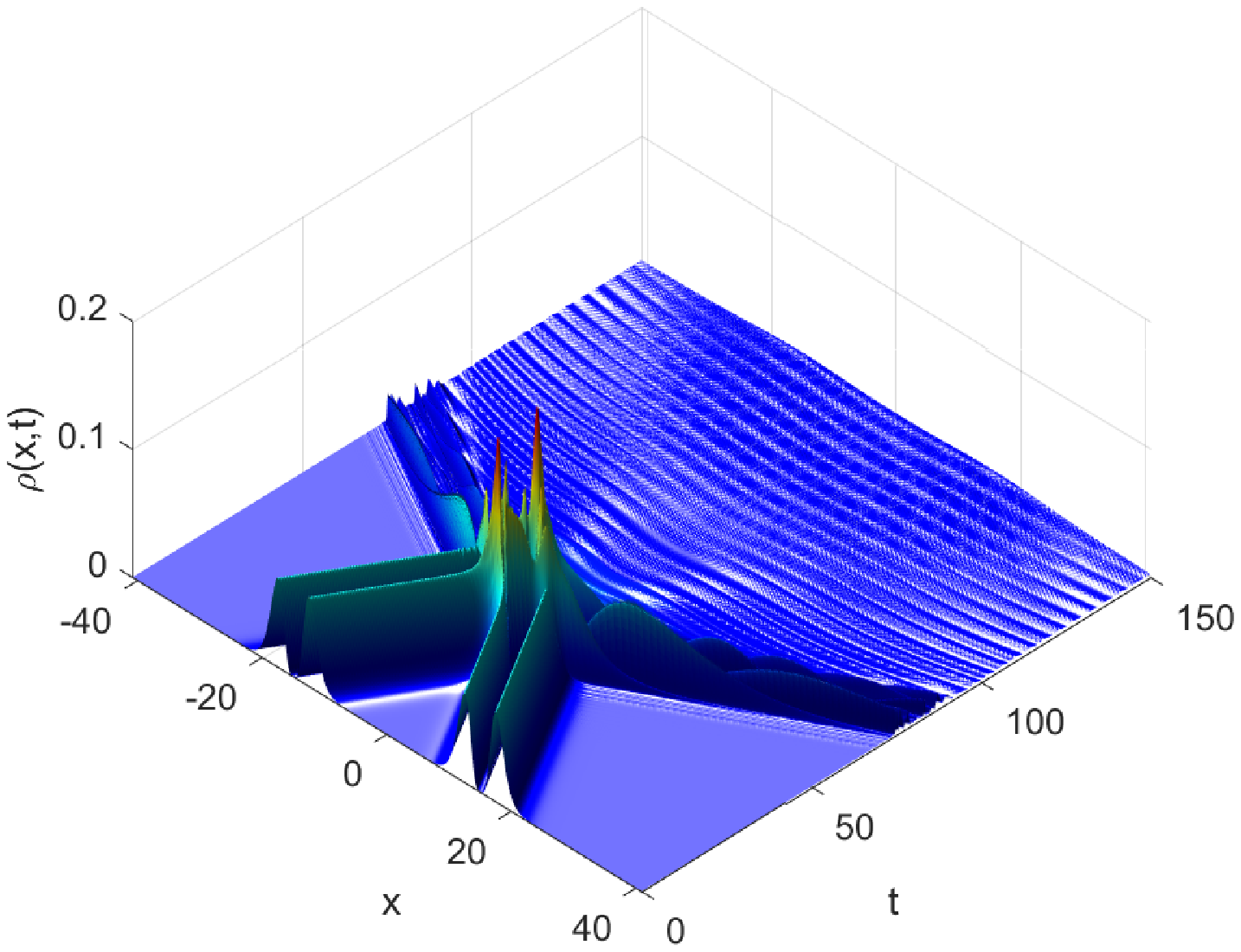}
\caption{Three-dimensional plots of the energy density of the two-kinks collision for the impact velocities $u=0.15$ (left) and $u=0.325$ (right). The remnant oscillon has a short lifetime. Here $p=3$.}
\end{figure}
\begin{figure}[htb]
\centering
\includegraphics[scale=0.5]{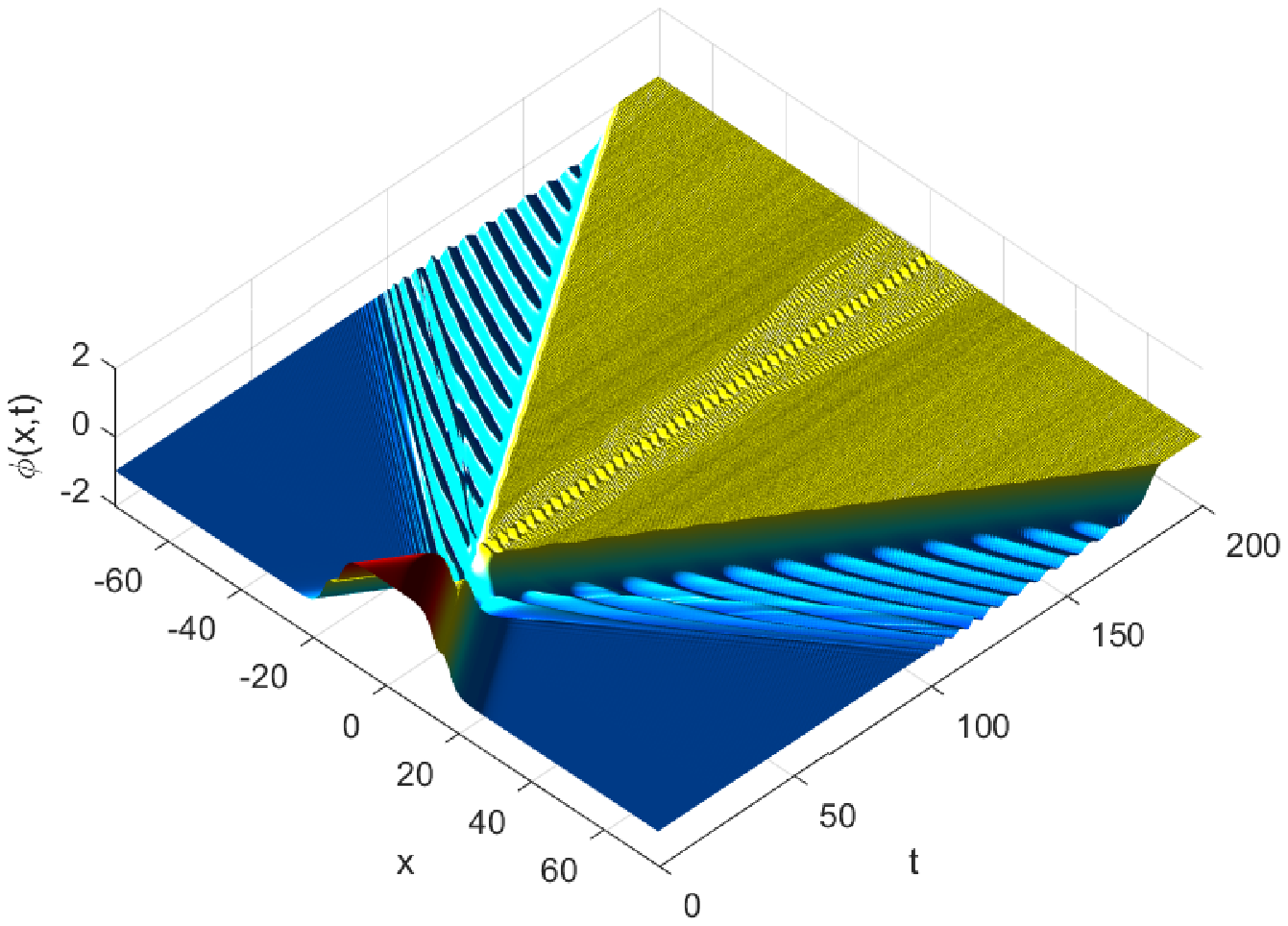}\includegraphics[scale=0.5]{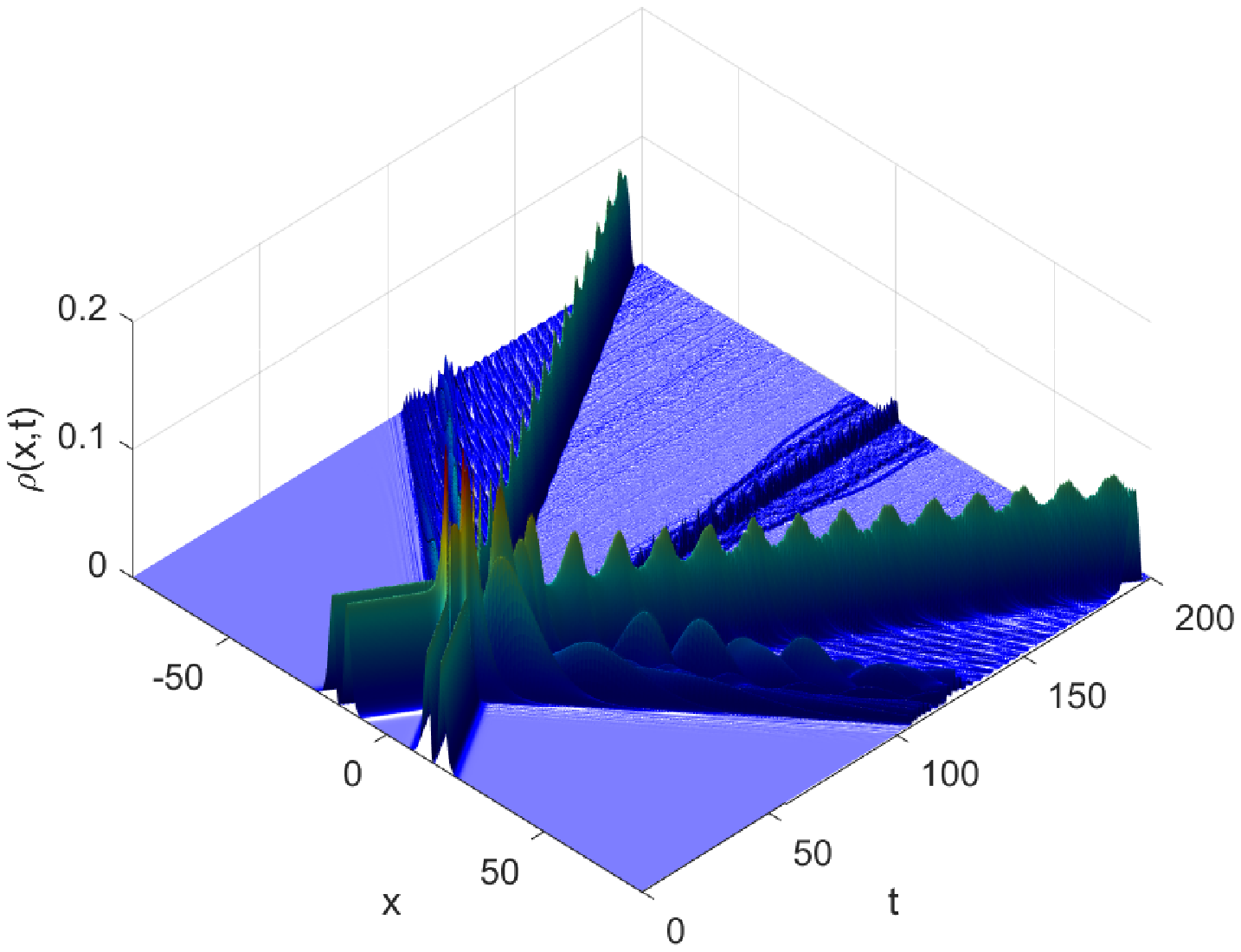}
\caption{Three-dimensional plots of the scalar field $\phi(x,t)$ and the energy density $\rho(x,t)$ for the impact velocity $u=0.5$ and $p=3$.} 
\end{figure}
\begin{figure}[htb]
\centering
\includegraphics[width=5.5cm,height=4.5cm]{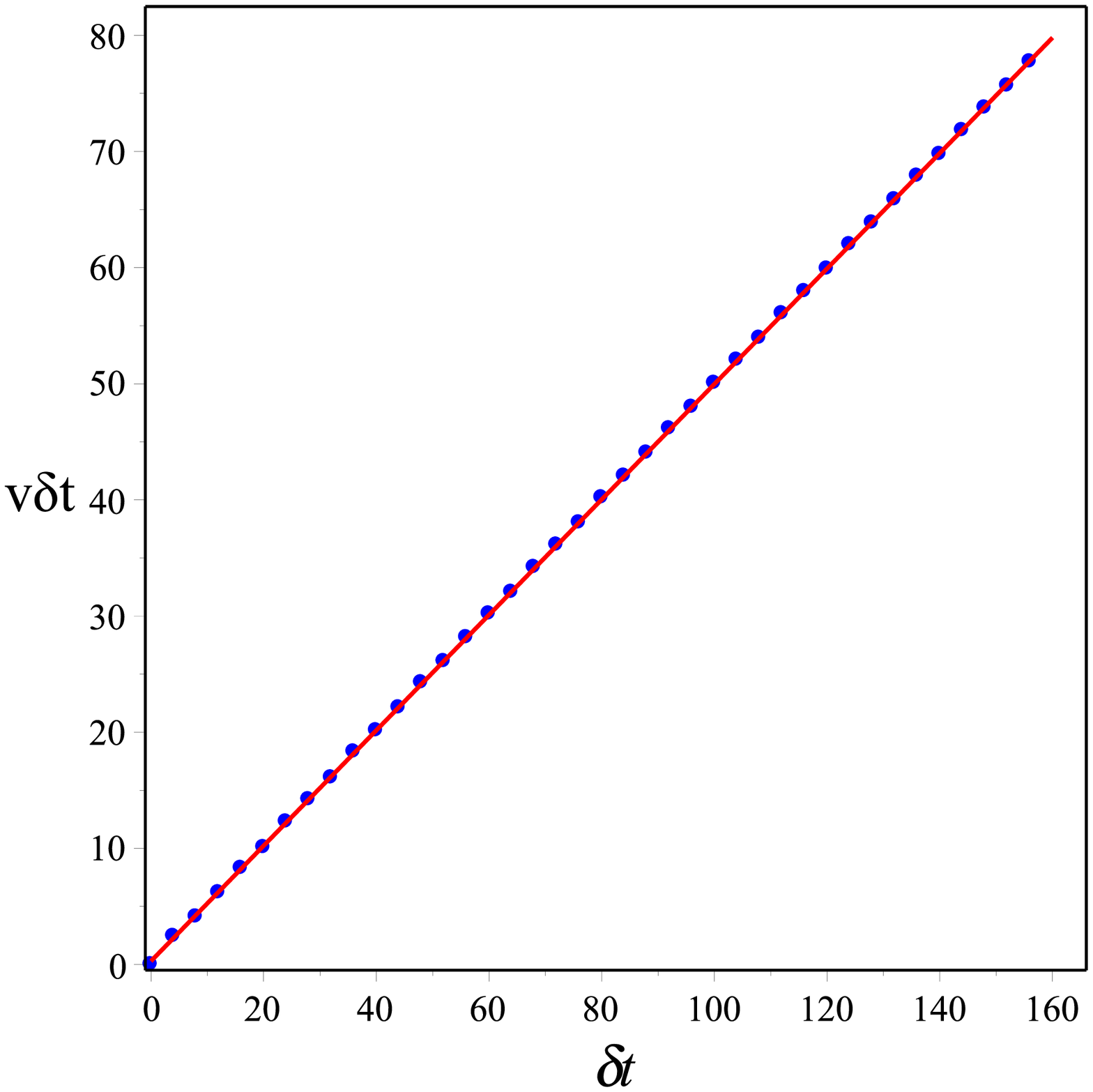}\hspace{0.5cm} 
\includegraphics[width=5.5cm,height=4.5cm]{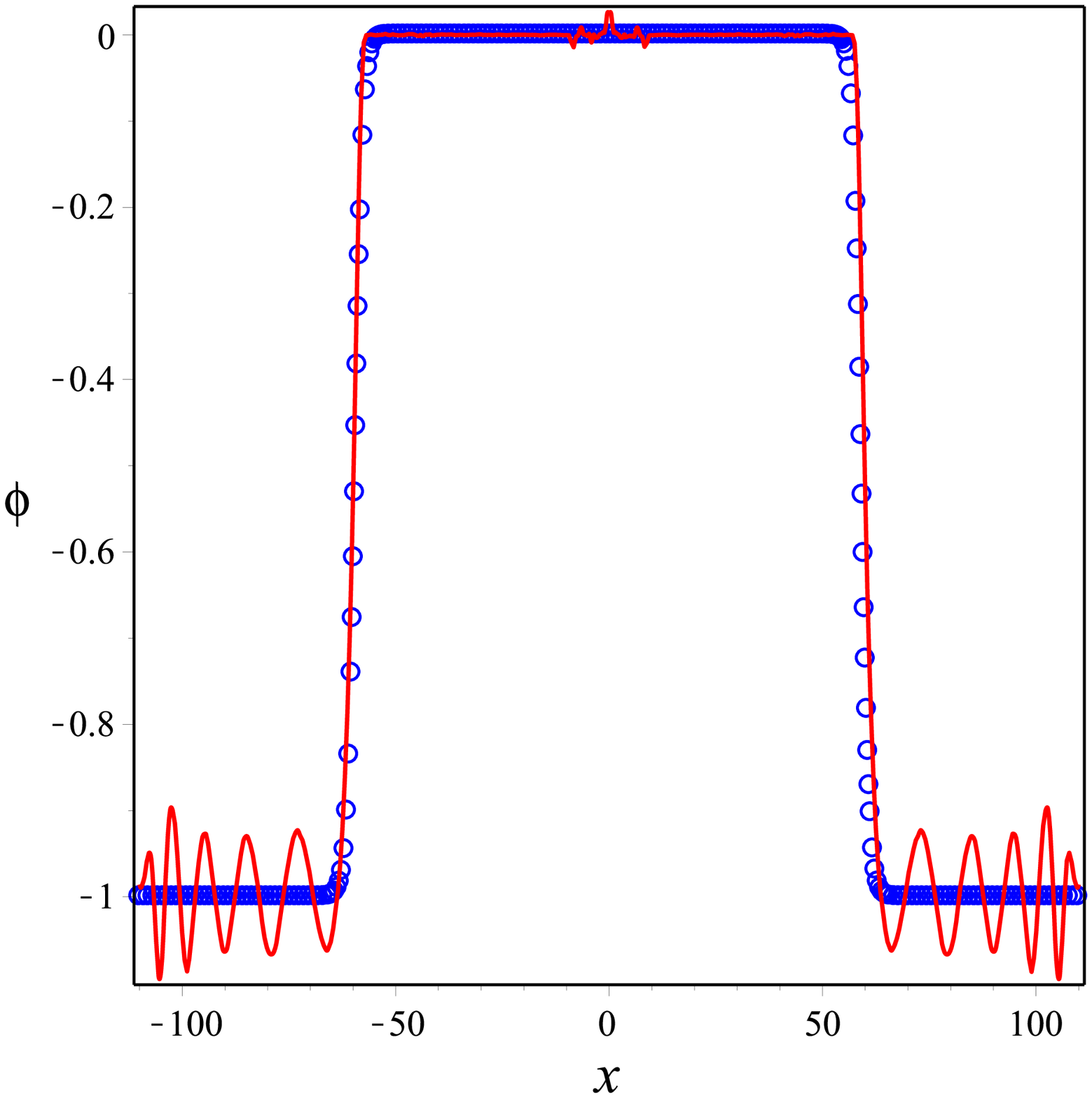} 
\caption{Plot of the position of one of the kinks $x$ versus $\delta t$. The straight line fits the sequence of the numerical locations (circles)  providing the escape velocity $v \approx 0.497$. In right panel we present the numerical (red line) and the exact (blue circles) profiles of the scalar field at $t=147$.}
\end{figure}

In general, the collision is highly inelastic with a considerable amount of radiated scalar field, contrary to the collision of kinks ($p=1$) where the process is quasi-elastic. As a consequence, both two-kinks are destroyed resulting in two main solutions. The first is the formation of a bound state or a remnant oscillon at the origin with a small amplitude and depending on the impact velocity the oscillon has a very short lifetime. We named this solution as the bound state and takes place if the impact velocity is smaller than a specific value or the limit velocity, $u_L$ \footnote{In the collision of kinks this velocity is called the critical velocity.}. This solution is illustrated in Fig. 1 with the plots of the energy density, $\rho(x,t)$

\begin{equation}
\rho(x,t) = \frac{1}{2}\left(\frac{\partial \phi}{\partial t}\right)^2 + \frac{1}{2}\left(\frac{\partial \phi}{\partial x}\right)^2 + V(\phi), \label{eq2.3}
\end{equation} 

\noindent for $p=3$ and impact velocities $u=0.15,0.325$.

In Ref. \cite{tiago} we have described the second solution occurring in general for $u>u_L$ as the formation and escape of oscillon-like structures together with a small oscillon at the origin. Fig. 2 shows the three-dimensional plots of the scalar field and the energy density for $p=3$ and impact velocity $u=0.5$. This figure corrects the Fig. 6 of Ref. \cite{tiago} where there is no bounce of the oscillon-like structures.

The novelty we present here is the identification of the moving oscillon-like structures as a pair of two-kinks moving apart to each other with an approximately constant velocity.   After inspecting the scalar field profiles in several instants, we have concluded that they are kinks associated with the following potential:

\begin{equation}
U(\varphi) = \frac{1}{2}\left(\kappa_0^2-(\varphi_0-\varphi)^2\right)^2, \label{eq2.4}
\end{equation}

\noindent where $\kappa_0$ and $\varphi_0$ are constants. We denote the moving kink/anti-kink solutions with speed $v$ by $\varphi_{K,\bar{K}}$ whose expressions are

\begin{equation}
\varphi_{K,\bar{K}}(x,t) = \varphi_0 \pm \kappa_0\tanh\left(\frac{\kappa_0(x-x_0-v t)}{\sqrt{1-v^2}}\right),\label{eq2.5}
\end{equation}

\noindent where $x_0$ denotes the initial position of the kink. The oscillon-like structures formed after the interaction of the two-kinks are described by 

\begin{equation}
\phi(x,t) = -1 - \kappa_0\tanh\left(\frac{\kappa_0(x-v \delta t)}{\sqrt{1-v^2}}\right)+
\kappa_0\tanh\left(\frac{\kappa_0(x+v \delta t)}{\sqrt{1-v^2}}\right). \label{eq2.6}
\end{equation}

\noindent Here $\delta t = t-t_{\mathrm{esc}}$ with $t_{\mathrm{esc}}$ being any time taken after the formation of the kinks.  The expression (\ref{eq2.6}) represents a pair of kinks moving in opposite directions with velocity $v$. We have found that \textit{all} emerging kinks are well fitted with $\kappa_0=0.5$ and $\phi_0=-0.5$ leaving the parameters $t_{\mathrm{esc}}$ and $v$ to be determined. Fig. 3 shows the plot of the position of one of the kinks versus $\delta t$ inferred from the energy density of Fig. 2, where the escape velocity is $v \approx 0.497$ with  $t_{\mathrm{esc}}=51$. Using these values in the exact expression (\ref{eq2.6}), we can reproduce the corresponding numerical profile (red line) of the scalar field at $t=147$ as shown in the right panel of Fig. 3 supporting the identification of the outcome as the escape of a pair of kinks. We emphasize that similar fittings can be done at any other instant $t > 51$. We call the present solution the escape of kinks.

The considerable amount of scalar field radiated away when a pair of two-kinks collide can be understood by inspecting the internal modes of a two-kink from which the energy can be stored and transferred. Recall that the internal modes of a two-kink is obtained after perturbing it linearly as $\phi(x,t)=\tanh^p(x/p)+\chi_n(x) \mathrm{e}^{i \omega_n t}$, and solving the resulting Schrodinger equation for the eigenfunctions $\chi_n(x)$ associated with the eigenfrequencies $\omega_n$. For the kink ($p=1$) the spectrum has a zeroth or translational mode, $\omega_0=0$, a discrete mode also known as the vibrational or shape-mode oscillations characterized by $\omega_{\mathrm{osc}}=\sqrt{3}$, and a continuous spectrum with frequencies $\omega_k=\sqrt{k^2+4} \geq 2$. In this case, the eigenfunctions correspond to dispersive waves that propagate to spatial infinity. The two-kinks have the same spectrum structure, but the discrete and the continuous modes have frequencies that depend on the parameter $p=1,3,5,..$ as shown in Table 1. Accordingly, the discrete frequency and the minimum frequency of the continuous spectrum, $\omega_{\mathrm{min}}$, diminishes if $p$ increases.  These results indicate that the spectrum of the radiation modes is broader for the two-kinks than the corresponding spectrum of the standard kinks ($p=1$) (see Table 1) where $\omega_{\mathrm{min}} =2$ for $p=1$ whereas for $\omega_{\mathrm{min}} \approx 0.6667$ for $p=3$. Therefore, the broader radiative spectrum of a two-kink might be the reason for the large fraction of radiated scalar field when a pair of two-kink interact. We also remark that the oscillatory component shown in Fig. 2  reveals the excitation of the vibrational mode.  

\begin{table}
	\centering
		\begin{tabular}{c|c|c}
		\hline
		$p$ & $\omega_{\mathrm{osc}}$ & $\omega_{\mathrm{min}}$ \\
		\hline
		\hline
		$1$ & $\sqrt{3} \approx 1.73205$ & 2.0\\
		\hline 
		$3$ & $0.54975$ & $0.6667$\\
	\hline
		$5$ & $0.33860$ & $0.4001$\\
	\hline
		$7$ & $0.243204$ &$0.2858$\\
	\hline
		\end{tabular}
		\caption{Approximate values of $\omega_{\mathrm{osc}}$ and $\omega_{\mathrm{min}}$ corresponding to $p=1,3,5,7$.}
\end{table}

\section{The critical solution, windows of escape and fractality} 

\begin{figure}[htb]
\centering
\includegraphics[scale=0.42]{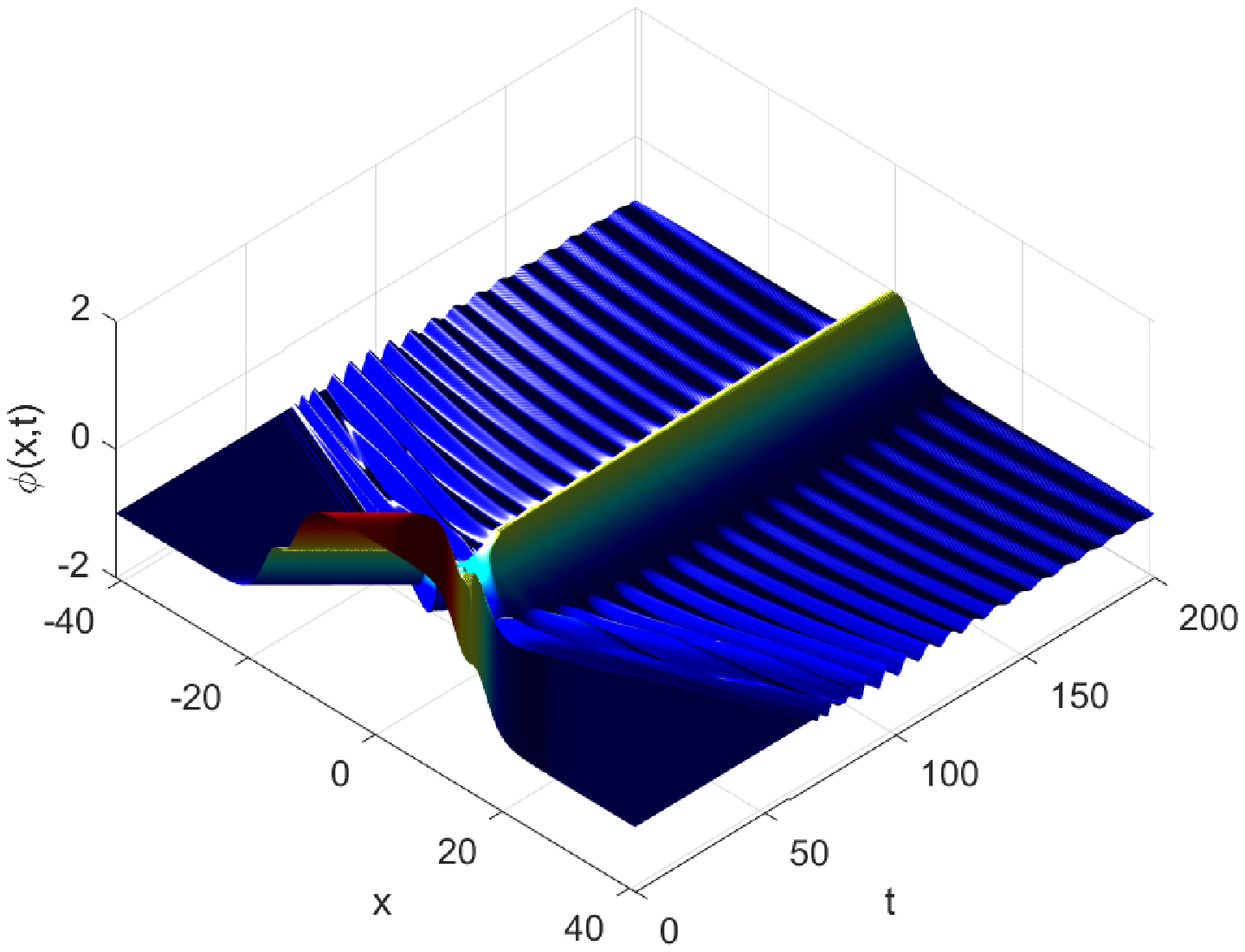}\includegraphics[scale=0.42]{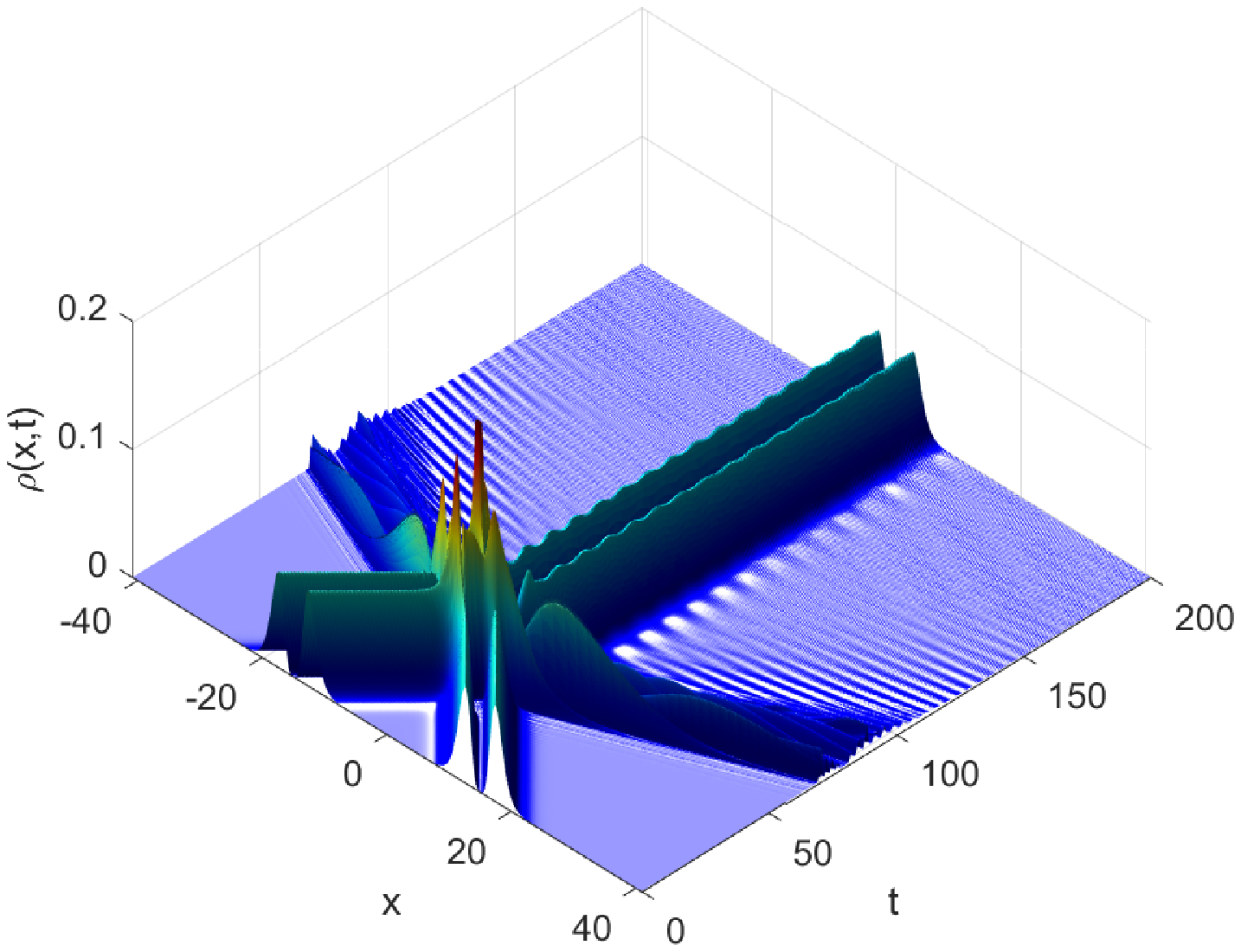}
\includegraphics[width=4.5cm,height=3.5cm]{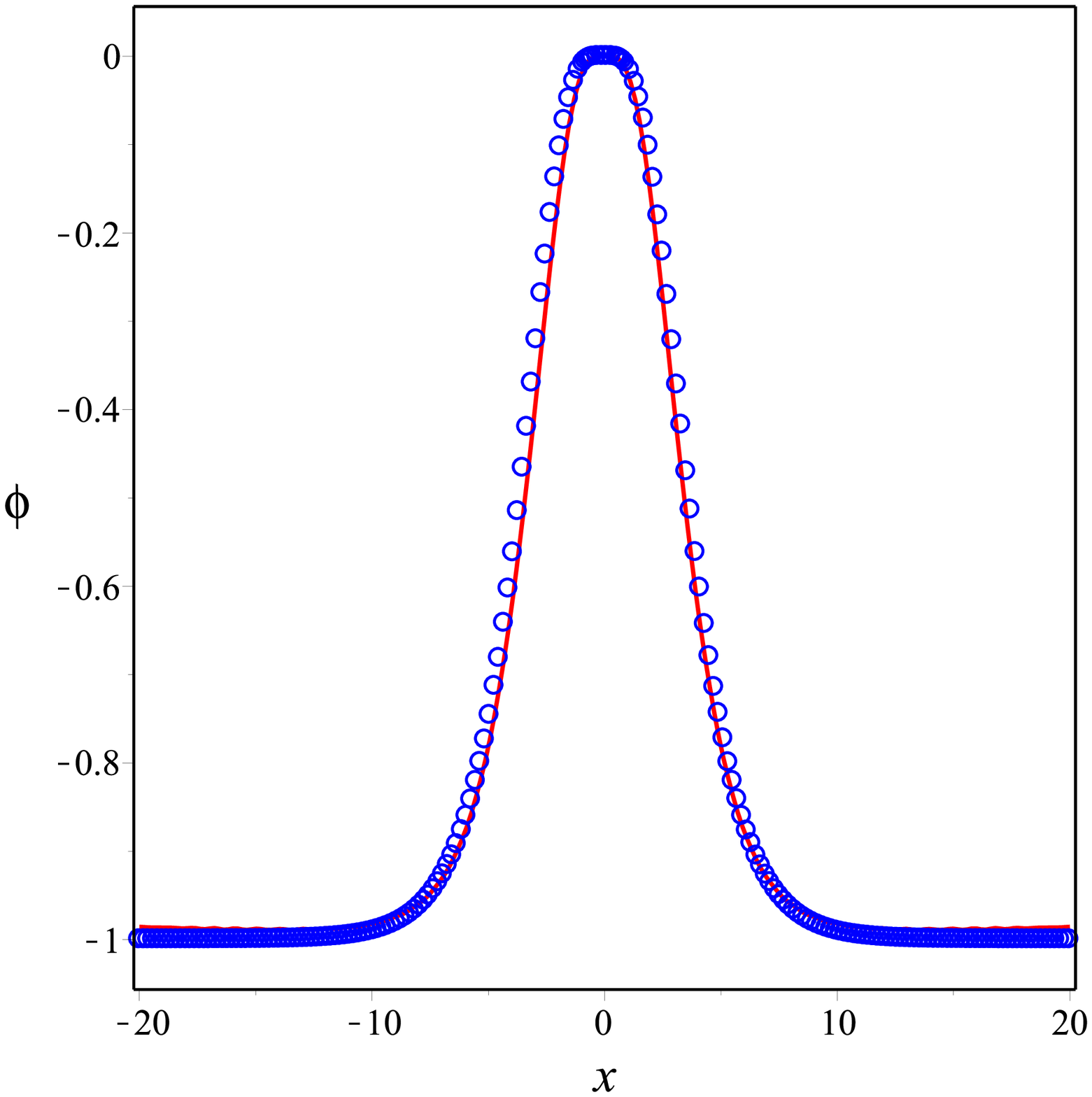}\hspace{3cm}
\includegraphics[width=4.5cm,height=3.5cm]{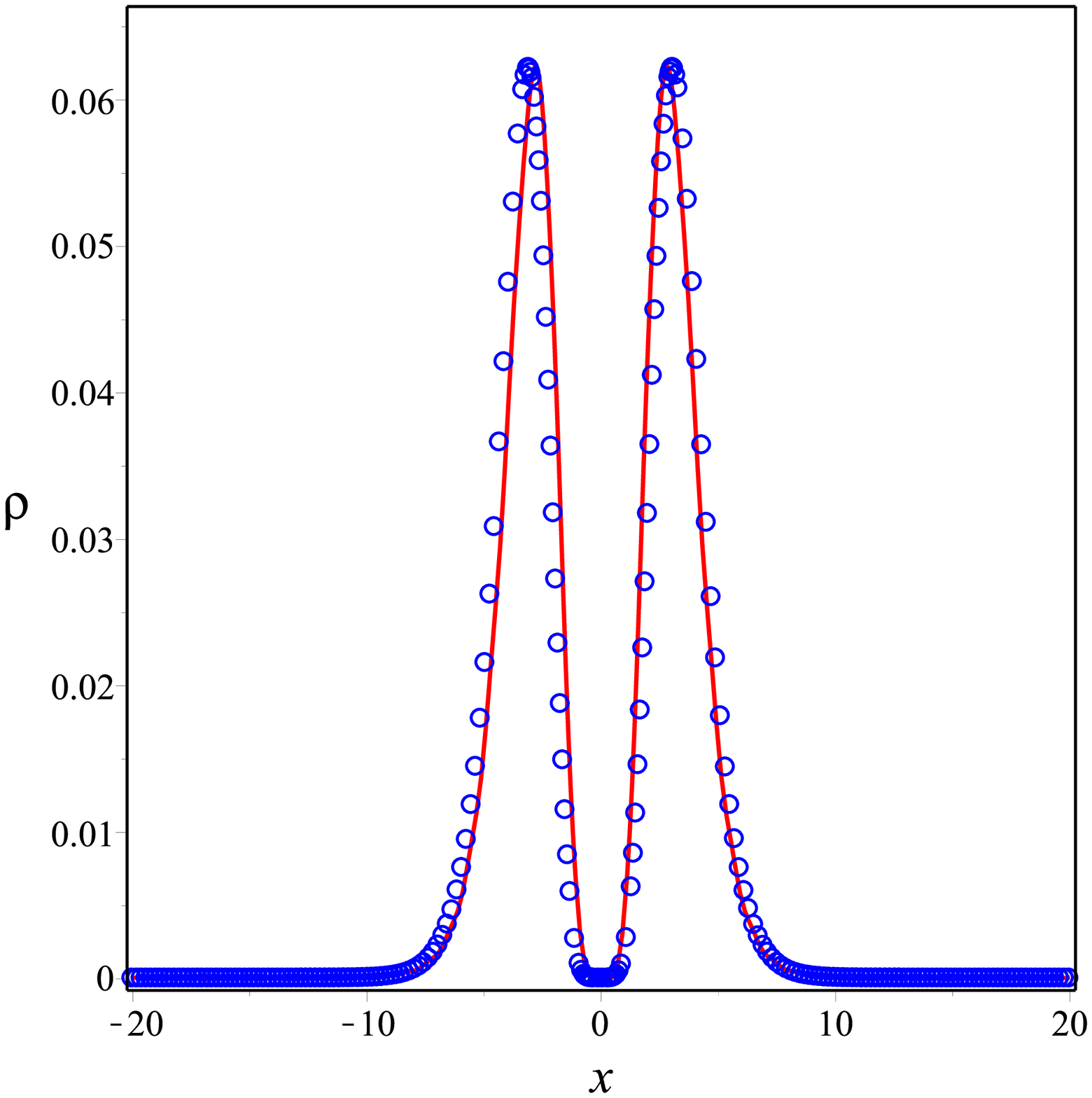}
\caption{Upper panels: critical configurations of the scalar field and the energy density for $p=3$ and generated with impact velocity $u=0.40553$. Lower panels: The numerical (line) and the exact (circles) profiles of the scalar field and the energy density at $t=200$. The fitting has the following parameters: $A_0 \simeq -1, b_0 \simeq 0.3353$ and $q=4$.}
\end{figure}
\begin{figure}[htb]
\centering
\includegraphics[scale=0.42]{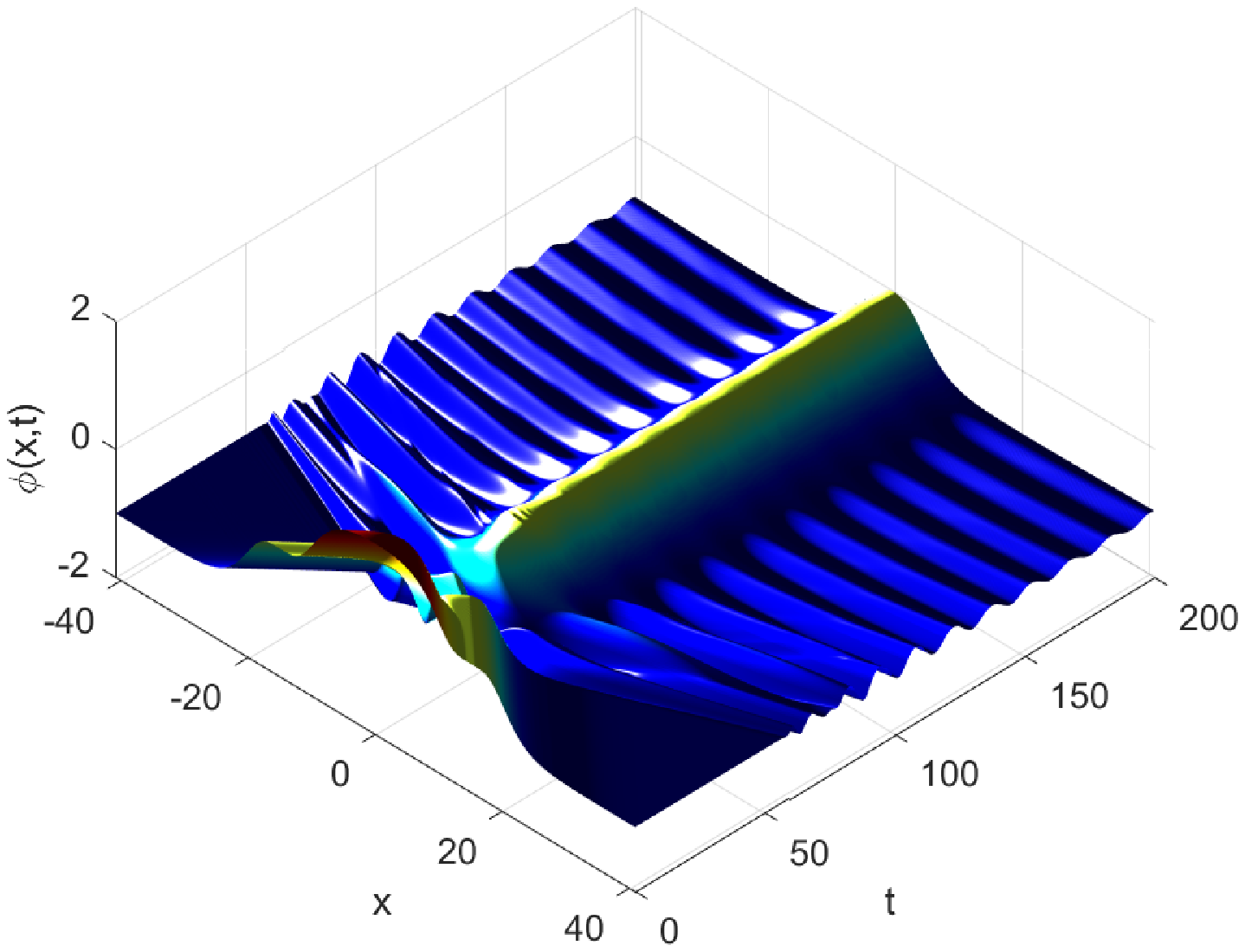}\includegraphics[scale=0.42]{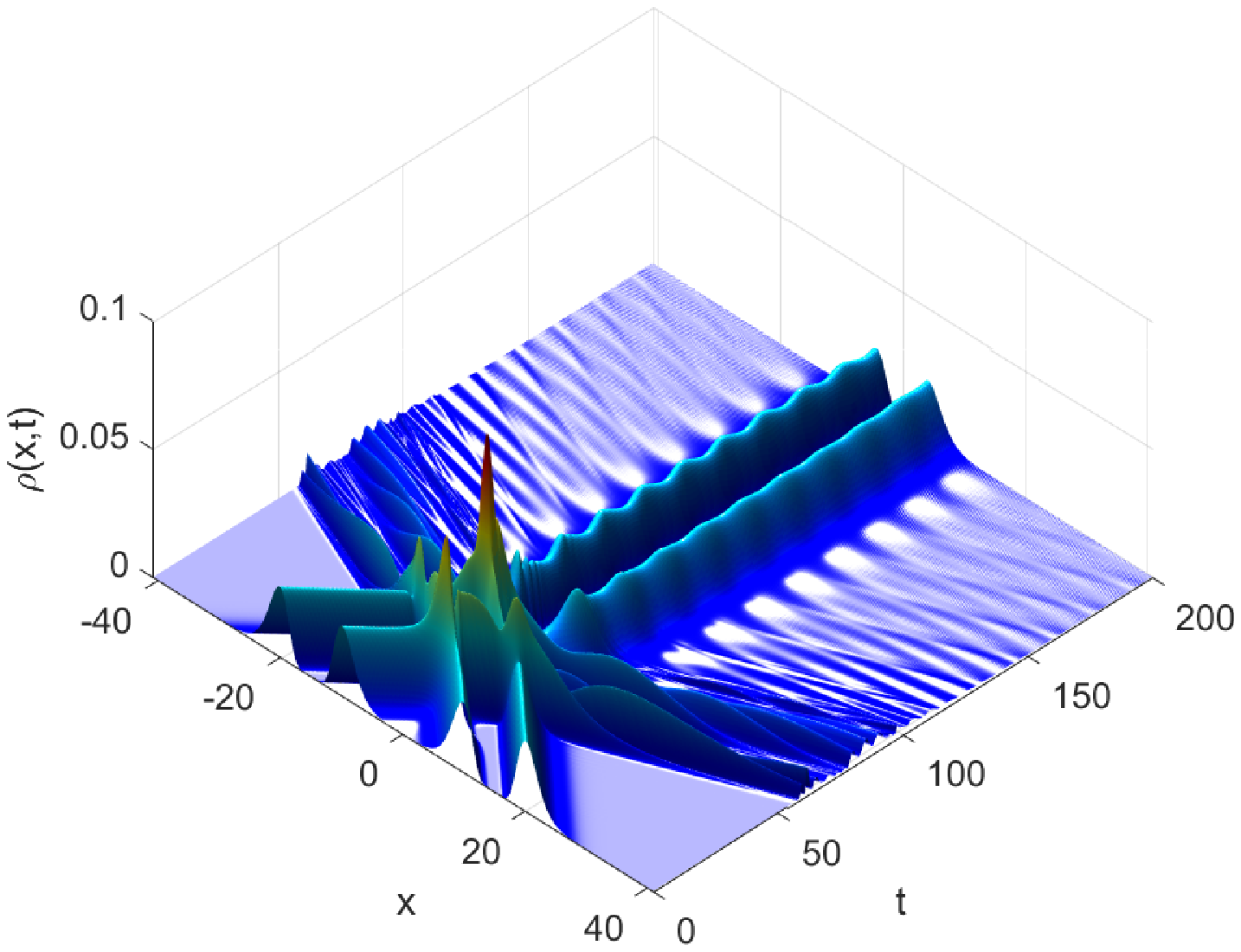}
\includegraphics[width=4.5cm,height=3.5cm]{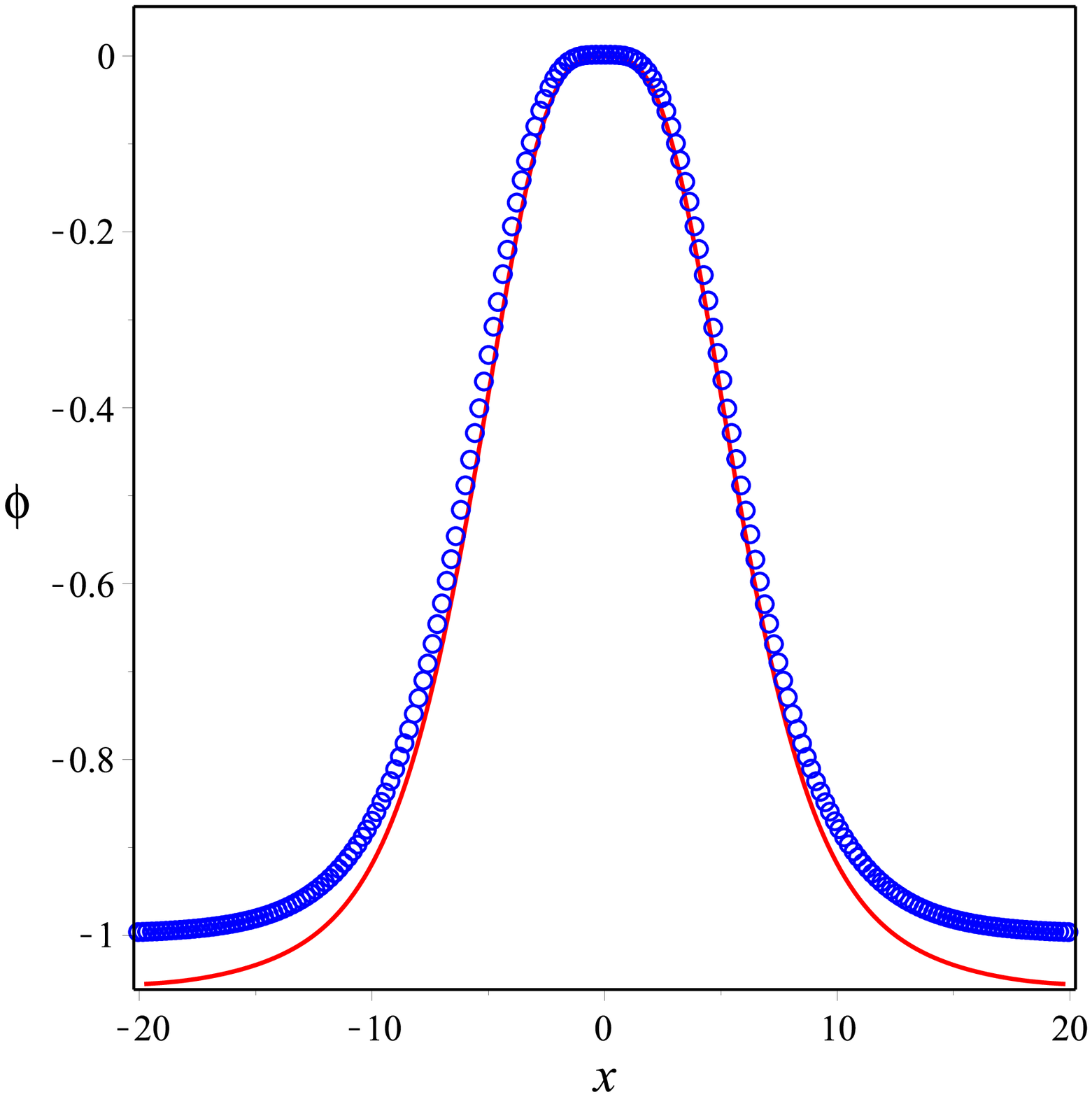}\hspace{3cm}\includegraphics[width=4.4cm,height=3.5cm]
{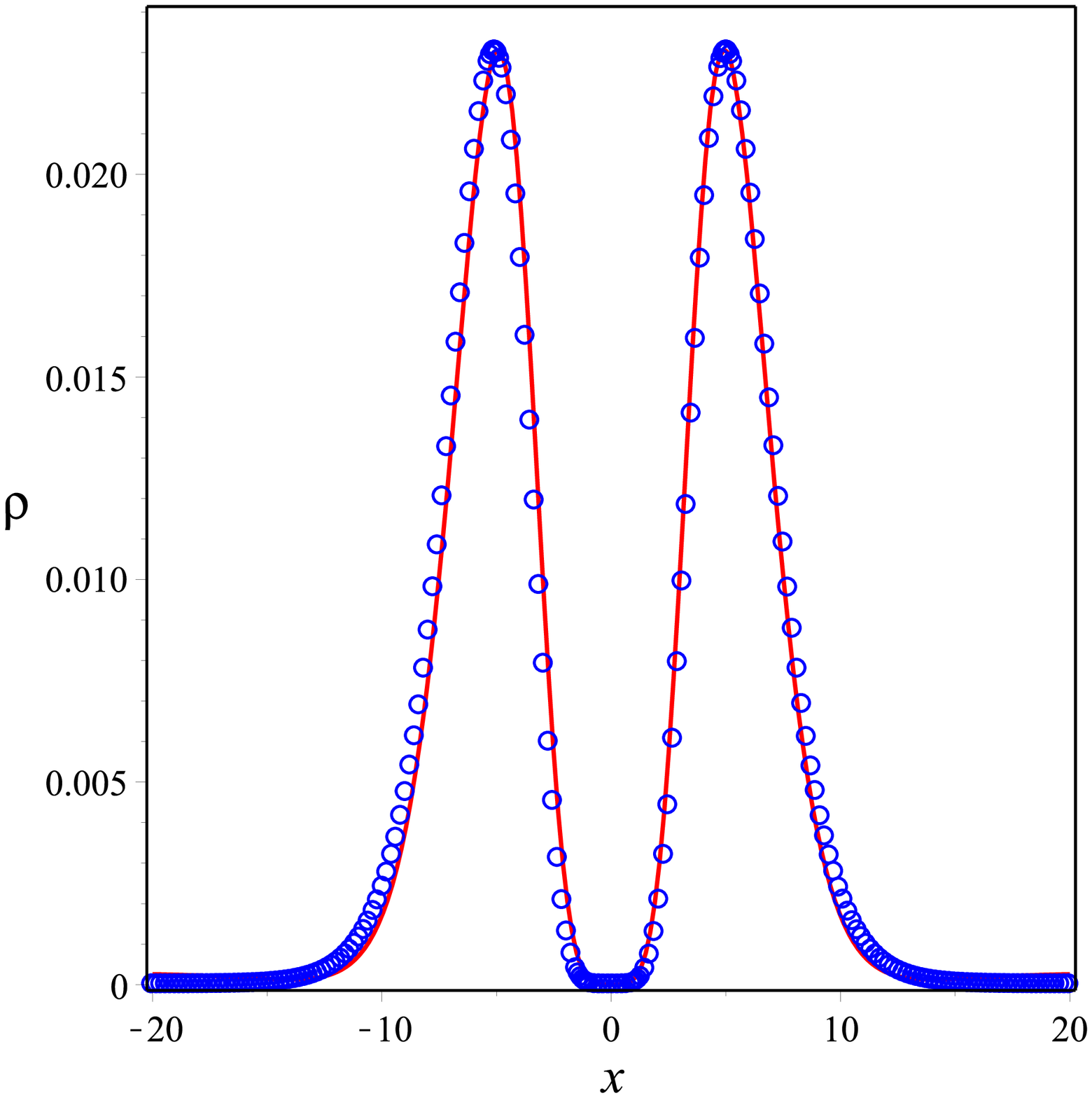}
\caption{Upper panels: critical configurations of the scalar field and the energy density for $p=5$ and generated with impact velocity $u=0.45004$. Lower panels: The numerical (line) and the exact (circles) profiles of the scalar field and the energy density at $t=125$. The fitting has the following parameters: $A_0 \simeq -1, b_0 \simeq 0.2041$ and $q=4$.}
\end{figure}
\begin{figure}[htb]
\centering
\includegraphics[scale=0.42]{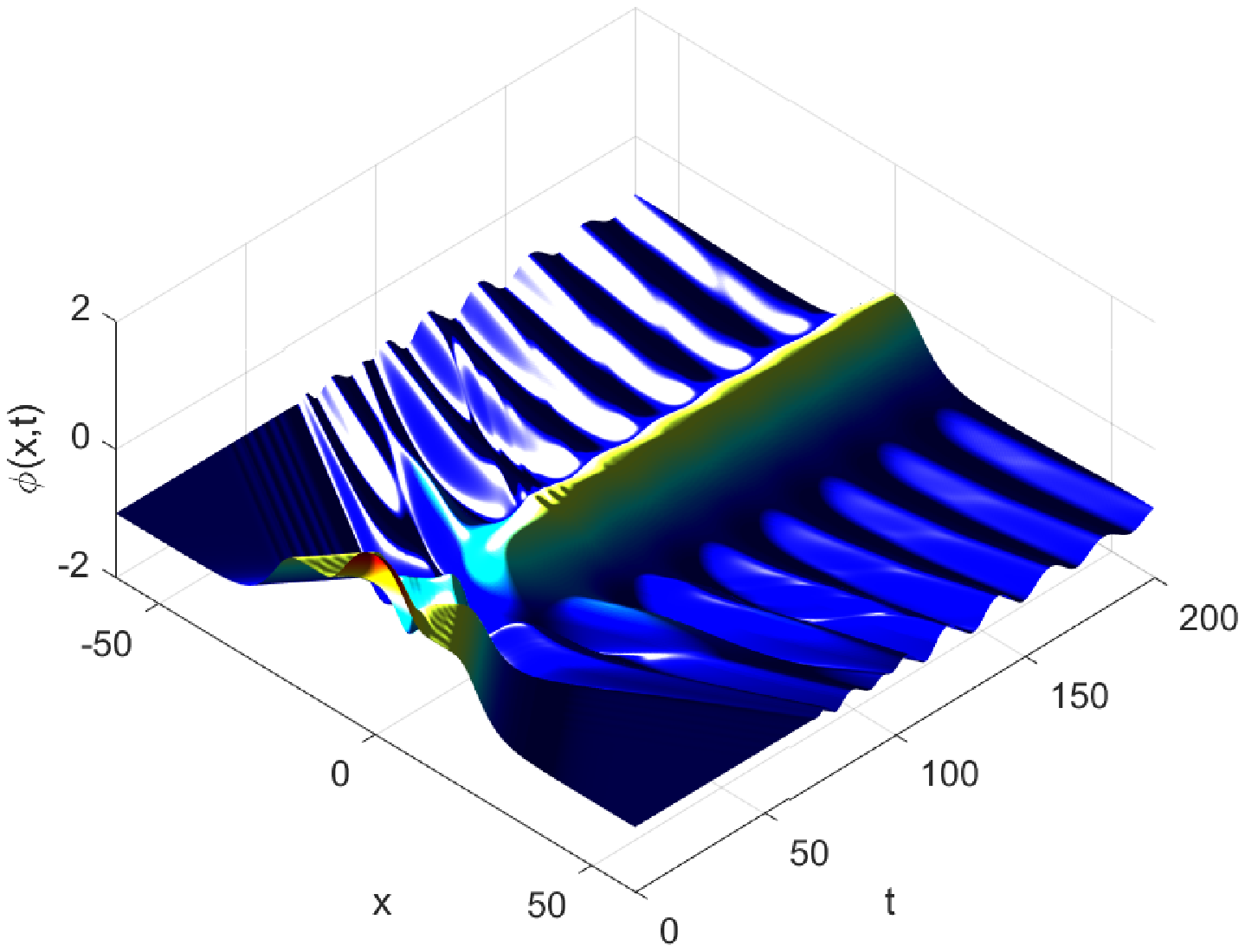}\includegraphics[scale=0.42]{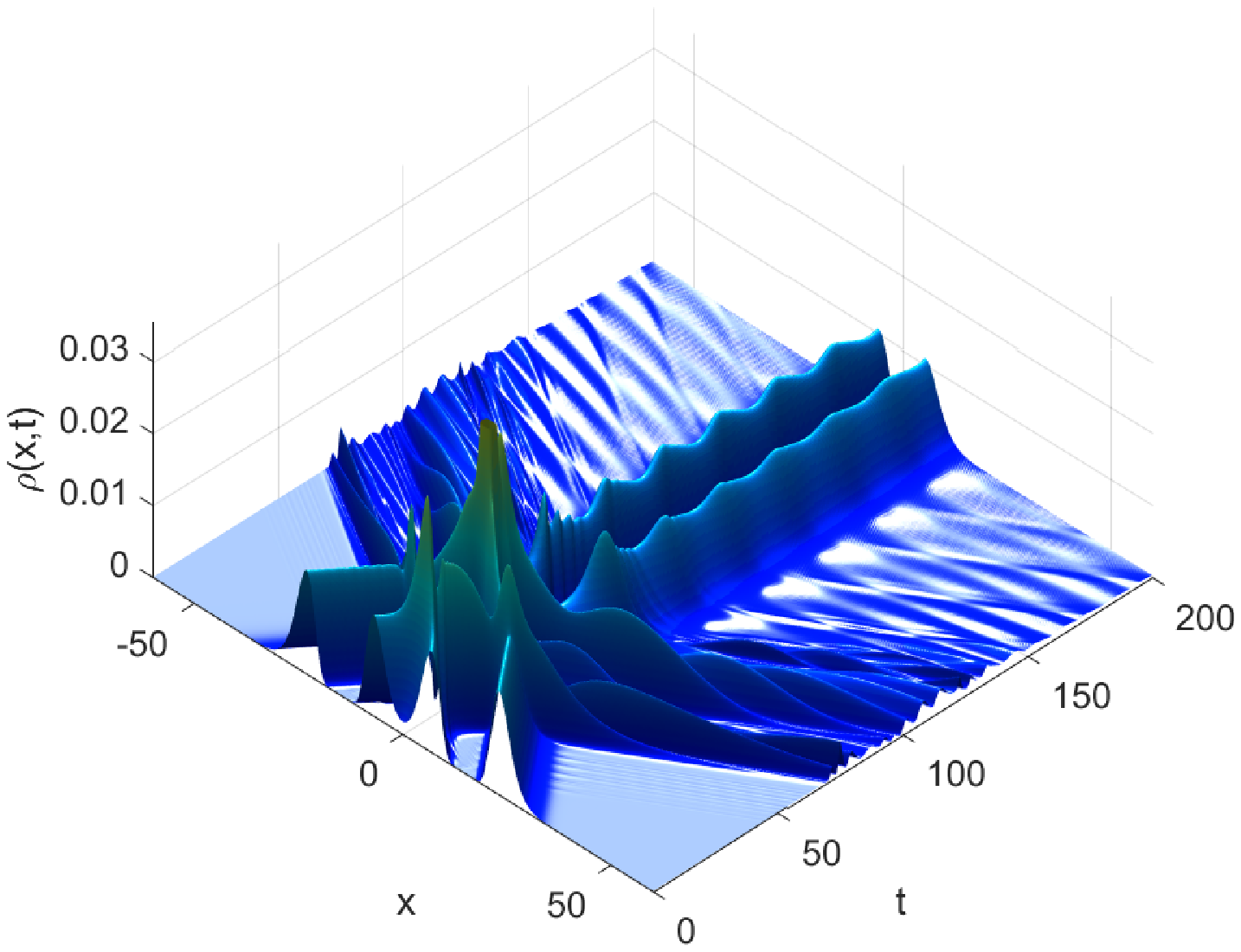}
\includegraphics[width=4.5cm,height=3.5cm]{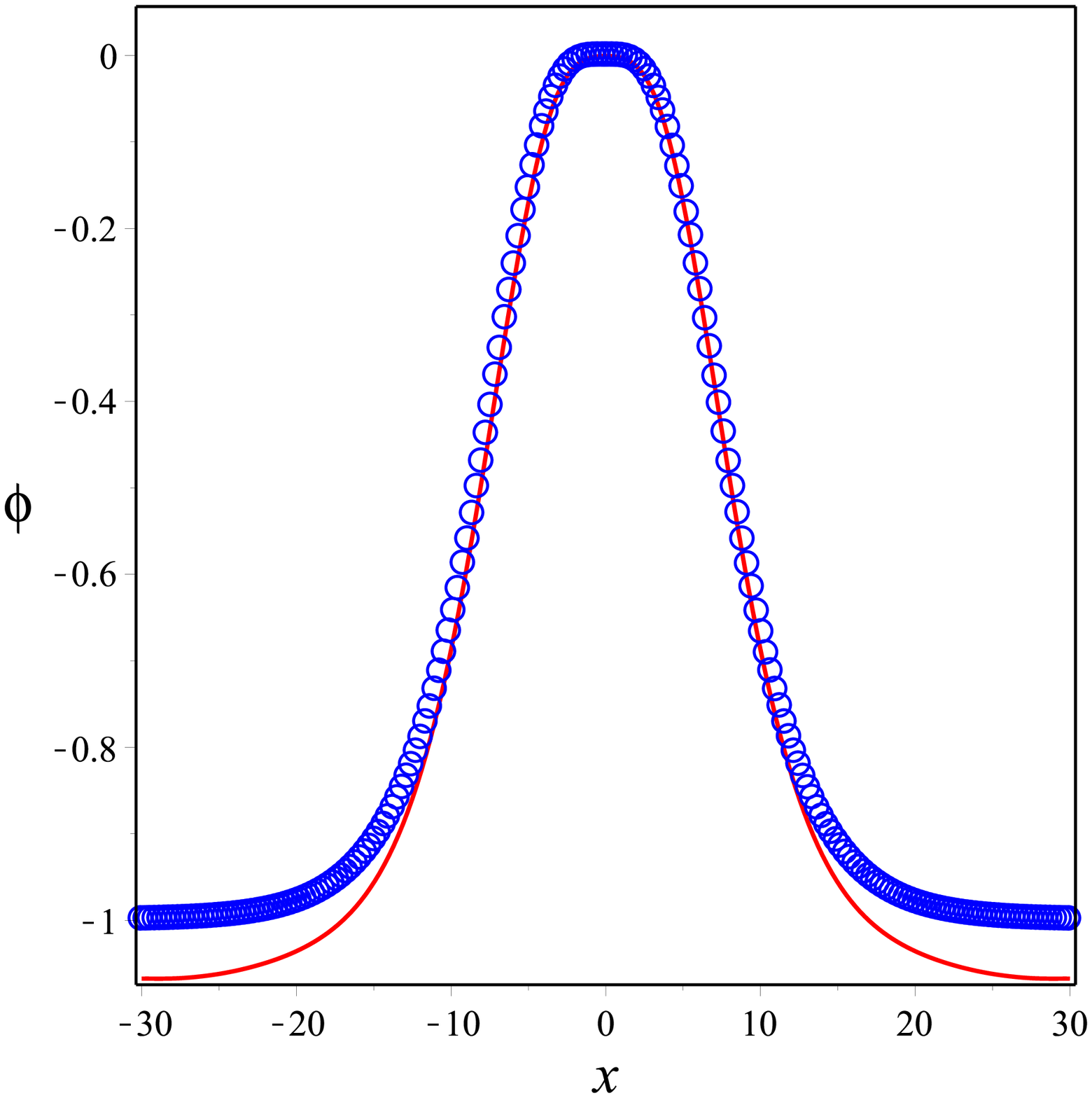}\hspace{3cm}\includegraphics[width=4.5cm,height=3.5cm]
{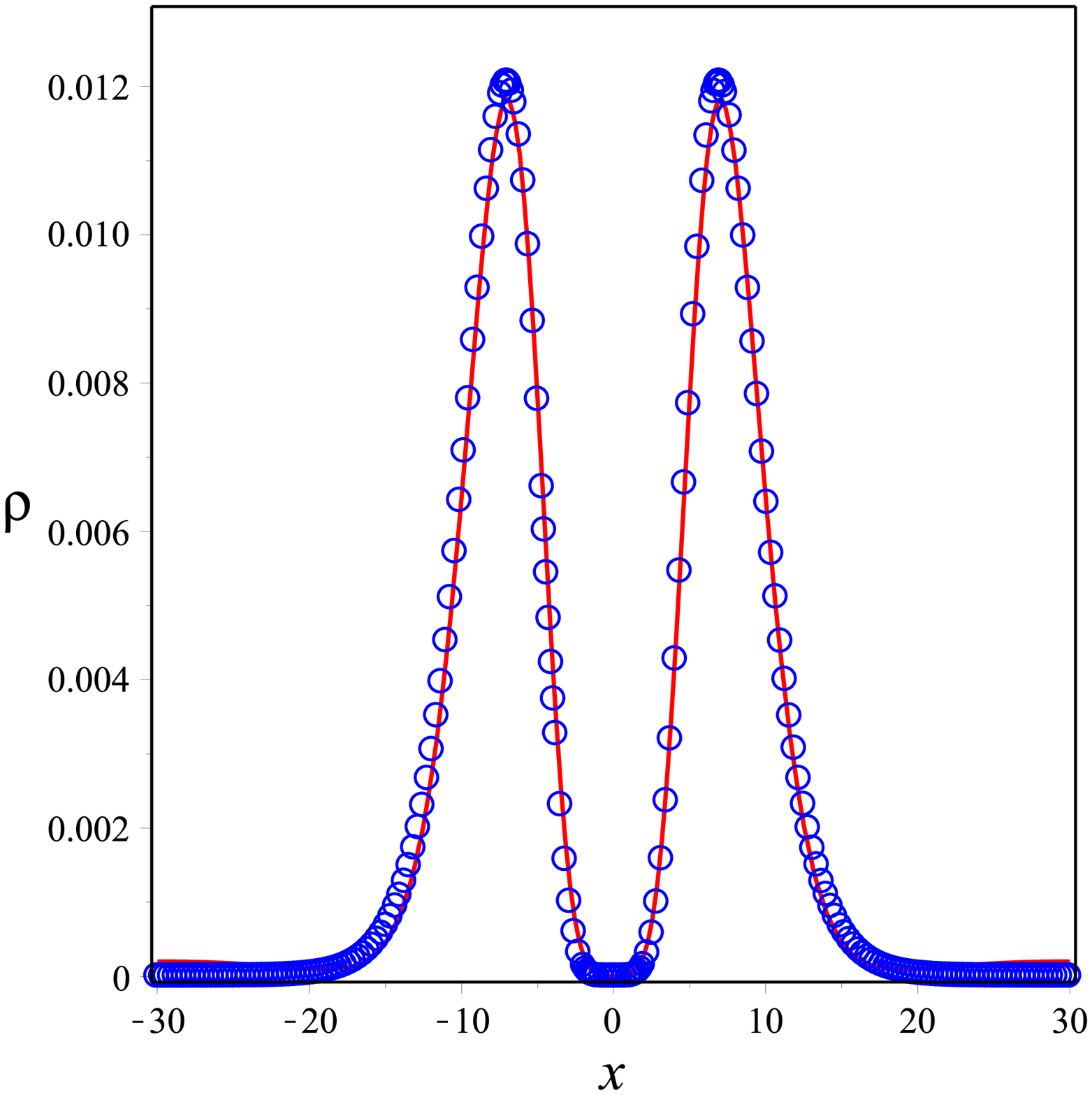}
\caption{Upper panels: critical configurations of the scalar field and the energy density for $p=7$ and generated with impact velocity $u=0.47647$. Lower panels: The numerical (line) and the exact (circles) profiles of the scalar field and the energy density at $t=162$. The fitting has the following parameters: $A_0 \simeq -1, b_0 \simeq 0.1477$ and $q=4$.}
\end{figure}
\begin{figure}[htb]
\centering
\includegraphics[width=12cm,height=1.2cm]{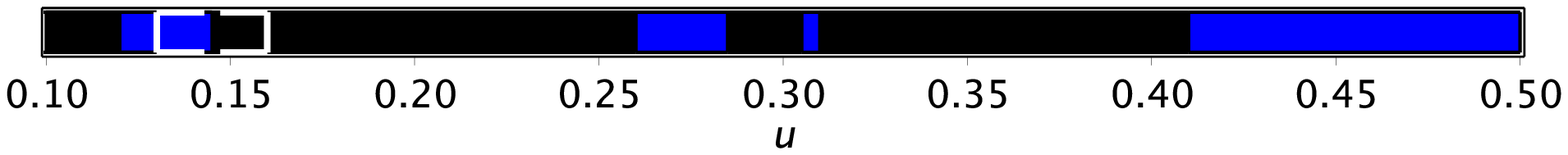}
\includegraphics[width=12cm,height=1.2cm]{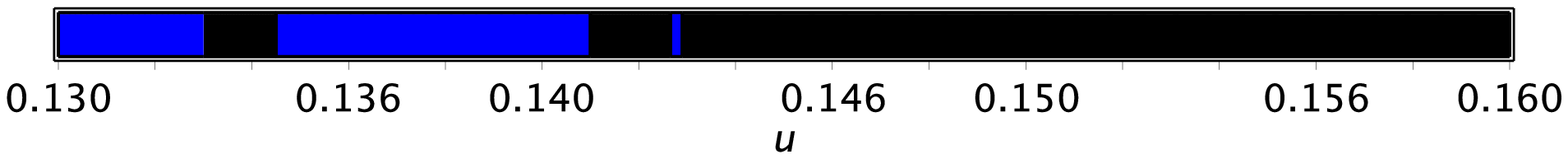}
\caption{Evidence that a boundary separating the solutions bound state (black) and escape of kinks (blue) is fractal. Here $p=3$.}
\end{figure}
\begin{figure}[htb]
\centering
\includegraphics[width=5.cm,height=4.5cm]{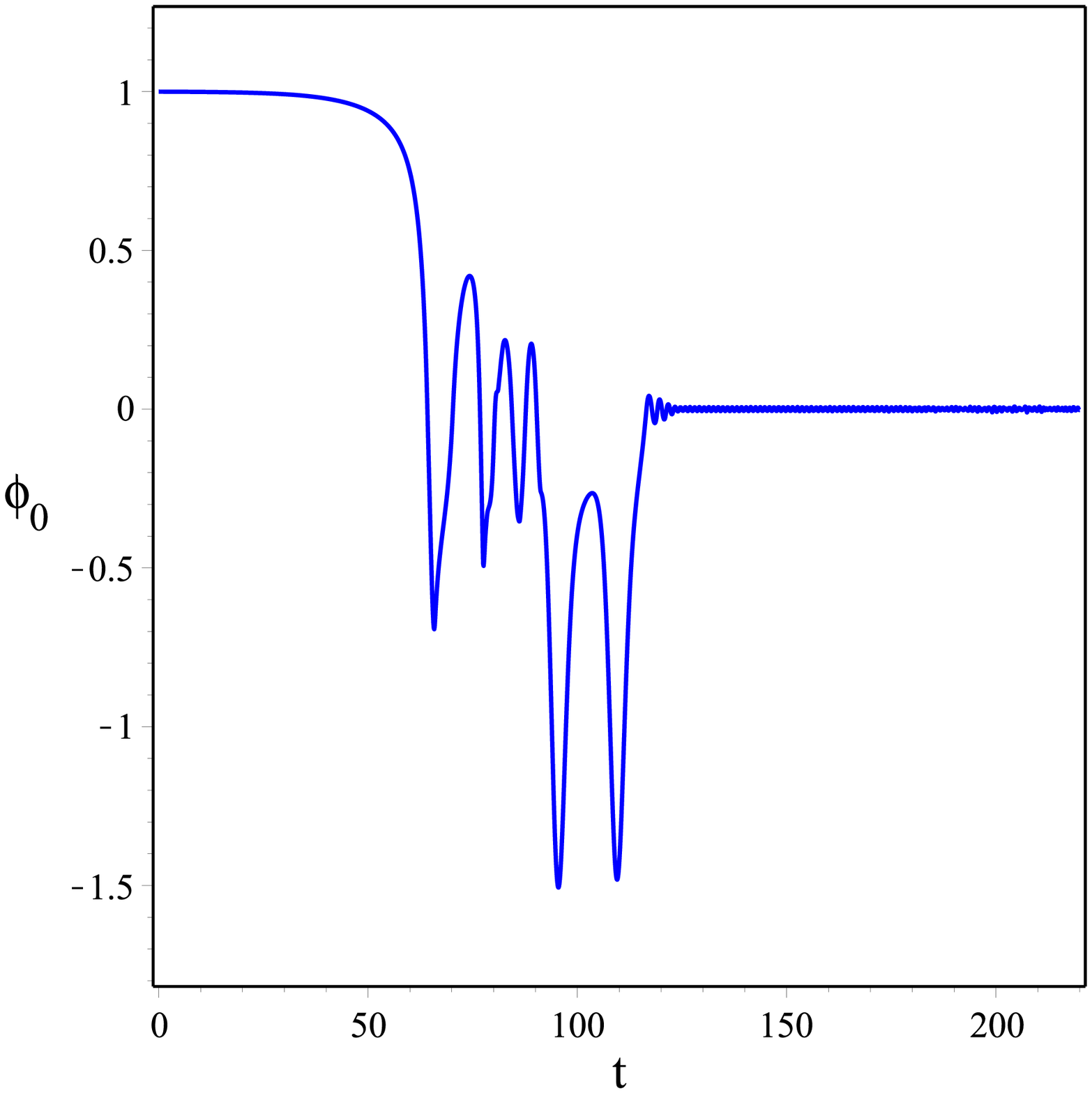}\includegraphics[width=5.cm,height=4.5cm]{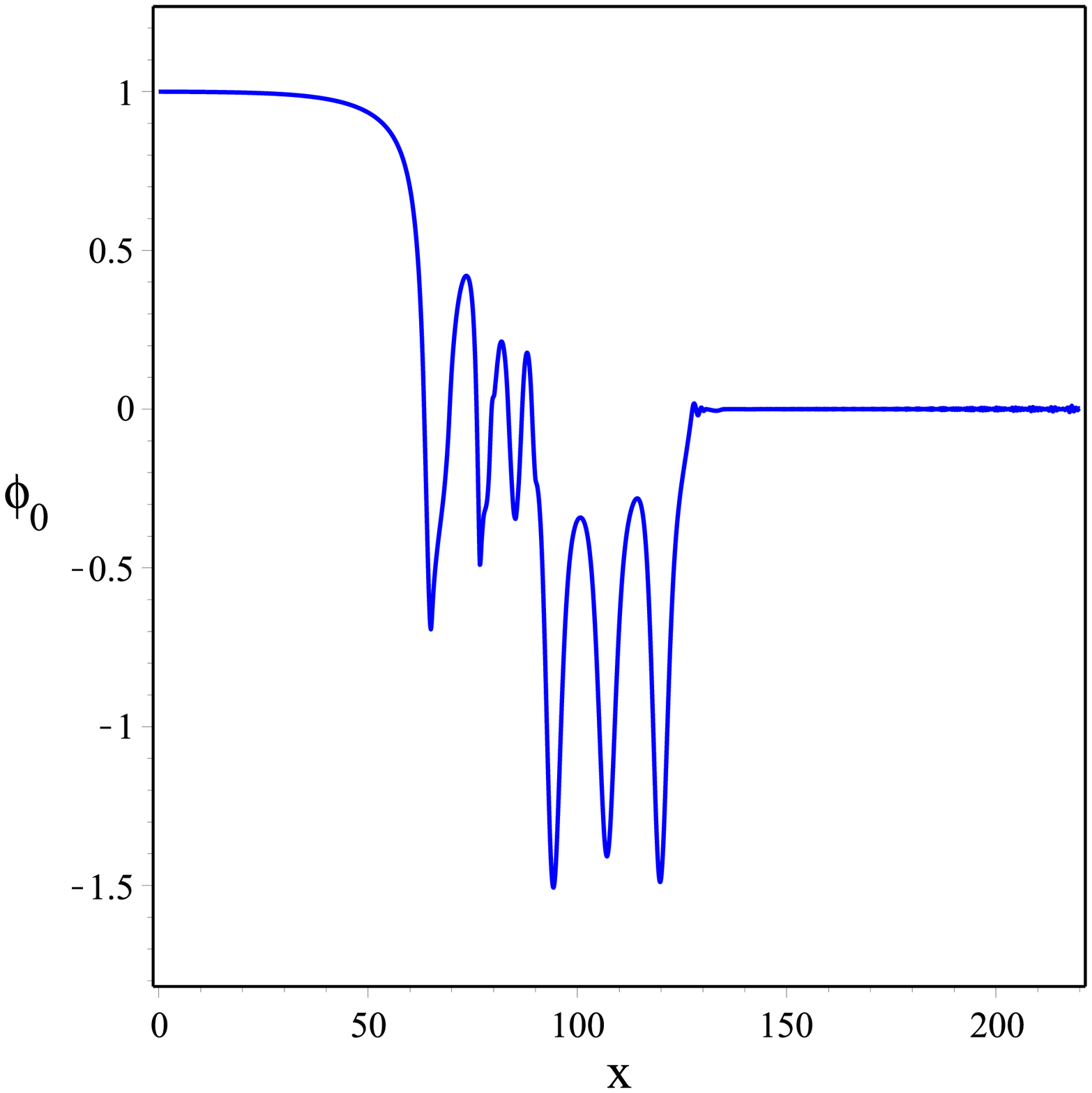}
\caption{Creation of kinks after one bounce obtained for $u=0.14050$ and after two-bounces for $u=0.14265$. The bounces occur about $t=100$.}
\end{figure}

The critical configurations constitute the third possible outcome resulting from the interaction of a pair of two-kinks and emerge if the impact velocity approaches to some specific values or the critical values. In Ref. \cite{tiago} we described the critical configurations by two oscillon-like structures that remain approximately at rest after moving apart from each other. These quasi-stationary structures were identified as nontopological lump-like defects described by \cite{tiago} 

\begin{equation}
\phi_{\mathrm{lump}}(x) =\phi_0+ A_0\tanh^q(b_0 x),\label{eq3.1}
\end{equation}

\noindent where $q$ is an even number, $ A_0, \phi_0$ and $b_0$ are constants. However, the numerical experiments indicate that for all critical configurations $\phi_0 \simeq 0$ and $A_0 \simeq -1$. In Figs. 4, 5 and 6 we present the three-dimensional plots of the scalar field and the energy density corresponding to the critical configurations for $p=3,5$ and $7$ respectively, and generated by the impact velocities $u=0.40553\,(p=3)$, $u=0.45004\,(p=5)$ and $u=0.47647\,(p=7)$. In the respective lower panels the numerical profiles (lines) of the scalar field and the energy density taken at $t=200\,(p=3)$, $t=125\,(p=5)$ and $t=162\,(p=7)$ are fitted with the lump-like solution (\ref{eq3.1}) with $q=4$ and $b_0=0.3353, 0.2041,0.1477$ for $p=3,5,7$ respectively. Despite the small oscillatory component present in the energy density, the fittings are very accurate.

The numerical experiments indicate a multiplicity of critical impact velocities. Another new feature we present here is that the presence of the critical solutions signalizes the windows of escape, i. e. intervals of the impact velocity below the limit velocity where the kinks described by Eqs. (\ref{eq2.5}) and (\ref{eq2.6}) emerge and escape. The collision of kinks \cite{kink_inter_1,kink_inter_2,kink_inter_3} also exhibits windows of escape known as the two-bounce windows related to the resonance between the kink solution and an internal mode that oscillates around it. In this case, the edges of the two-bounce are fractal containing a hierarchical structure of $n \geq 3$ bounce windows. For the collision of two-kinks, we have found evidence that the boundaries of the windows of escape have a fractal structure after zooming a region containing a bound state (black) and escape (blue) as shown in Fig. 7. Moreover, there is also an evidence of a similar hierarchical structure of the bounce windows as illustrated by the plots of the scalar field at the origin displaying the two and three bounces before the escape of the kinks (see Fig. 8)

\section{Lump-like defects as metastable configurations}%

There is no doubt that identifying nontopological lump-like defects as the critical configurations emerging from the interaction of two-kinks is an unexpected result. These structures are unstable under small perturbations and therefore unlikely to be formed after a dynamical process, but the last finding suggests that lump-like defects might develop a metastable configuration. 

We consider here a numerical experiment aimed to produce critical configurations by perturbing a lump-like defect conveniently and therefore to confirm its metastabilty character. We evolved a lump-like defect  excited by a perturbation according to the following initial data: 

\begin{equation}
\phi(0,x) = -\tanh^p\left(\frac{x}{p}\right) + \delta_+(x) + \delta _-(x),\label{eq4.1}
\end{equation}

\noindent where $p$ is an even number and 

\begin{equation}
\delta_{\pm}(x) =-B_0\mathrm{e}^{-(x \pm \bar{x}_0)^2/\sigma^2}.\label{eq4.2}
\end{equation}

\noindent Here $B_0$ is a free parameter representing the amplitude of the perturbations placed symmetrically at $x=\pm \bar{x}_0 = \pm 20$ and $\sigma=1.0$ (see Fig. 9). We have evolved the above initial data choosing $p=4$, but we also examined other even values of $p$ obtaining similar results. In general, as the system evolves a fraction of the perturbation escapes and another is attracted to the lump-like defect situated at the origin. For small values of $B_0$, the perturbation eventually breaks up the lump-like defect causing the emission of part of the scalar field leaving a bound state at the origin. We present this case in Fig. 9 with the evolution of the scalar field and the energy density for $B_0=0.1$.


After varying the parameter $B_0$, we have found some special values of $B_0$, the critical values, whose approach produces, after a period of an intricate dynamics, metastable structures that we have identified as nontopological lump-like defects described by Eq. (\ref{eq3.1}). We illustrate in Fig. 10 the formation of a lump-like defect after setting $B_0=0.2649705$. The defect is formed at $t \approx 60$ and remain unaltered until $t \approx 150$. We provided the fitting of the scalar field and the energy density using Eq. (\ref{eq3.1}) at $t = 150$. Therefore, we have found strong numerical evidence that unstable lump-like defects can emerge as the result of perturbating another lump-like defect conveniently. In this instance, we can understand that the perturbation described by Eq. (\ref{eq4.2}) reproduces the later stages of the collision of a pair of two-kinks defects with the parameter $B_0$ playing the role of the impact velocity $u$. As such, approaching it to one of the critical values gives rises the formation of a lump-like defect described by Eq. (\ref{eq3.1}).

\begin{figure}[htb]
\centering
\includegraphics[width=4.5cm,height=4.cm]{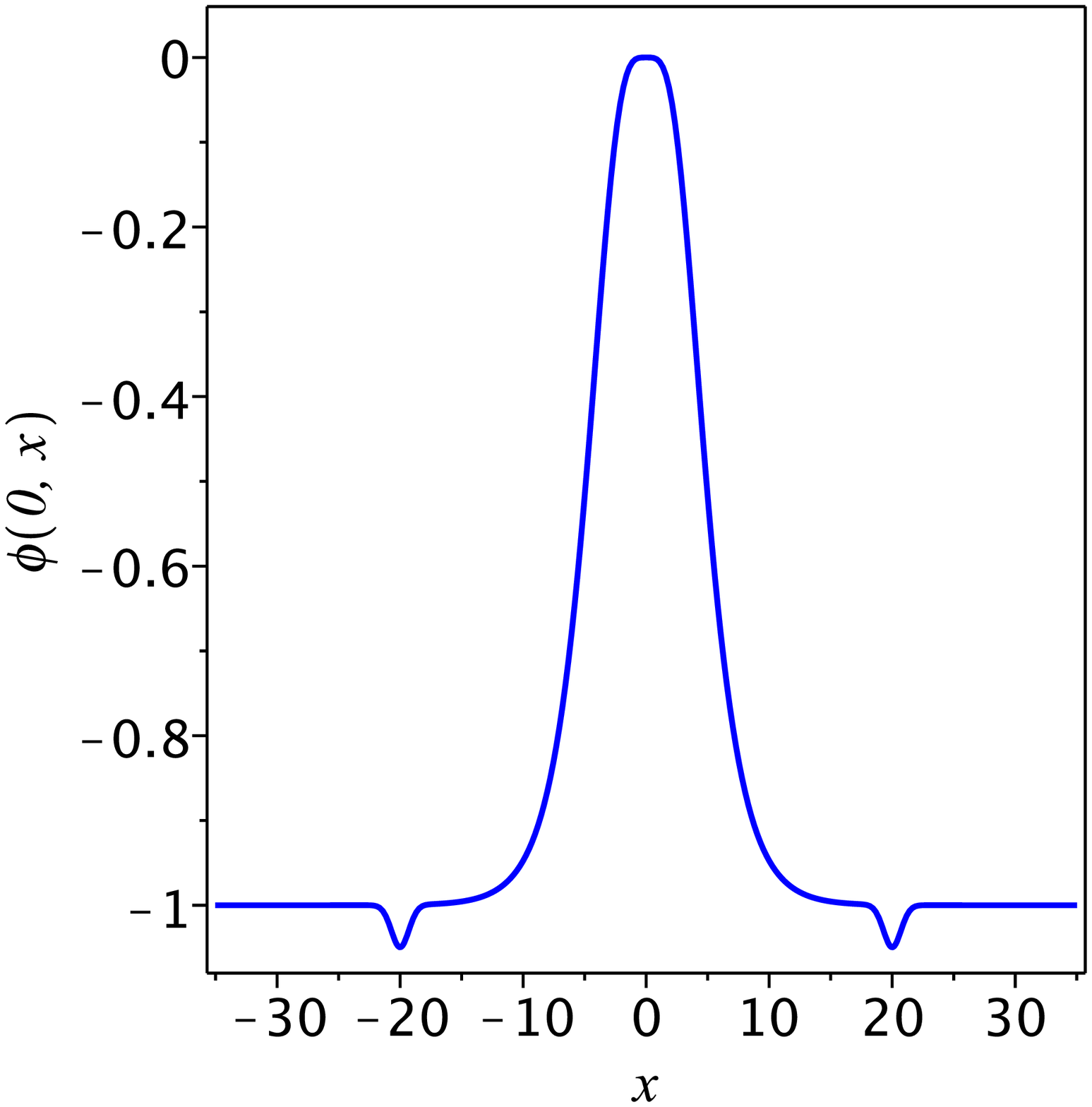} \hspace{2cm}
\includegraphics[width=4.5cm,height=4.cm]{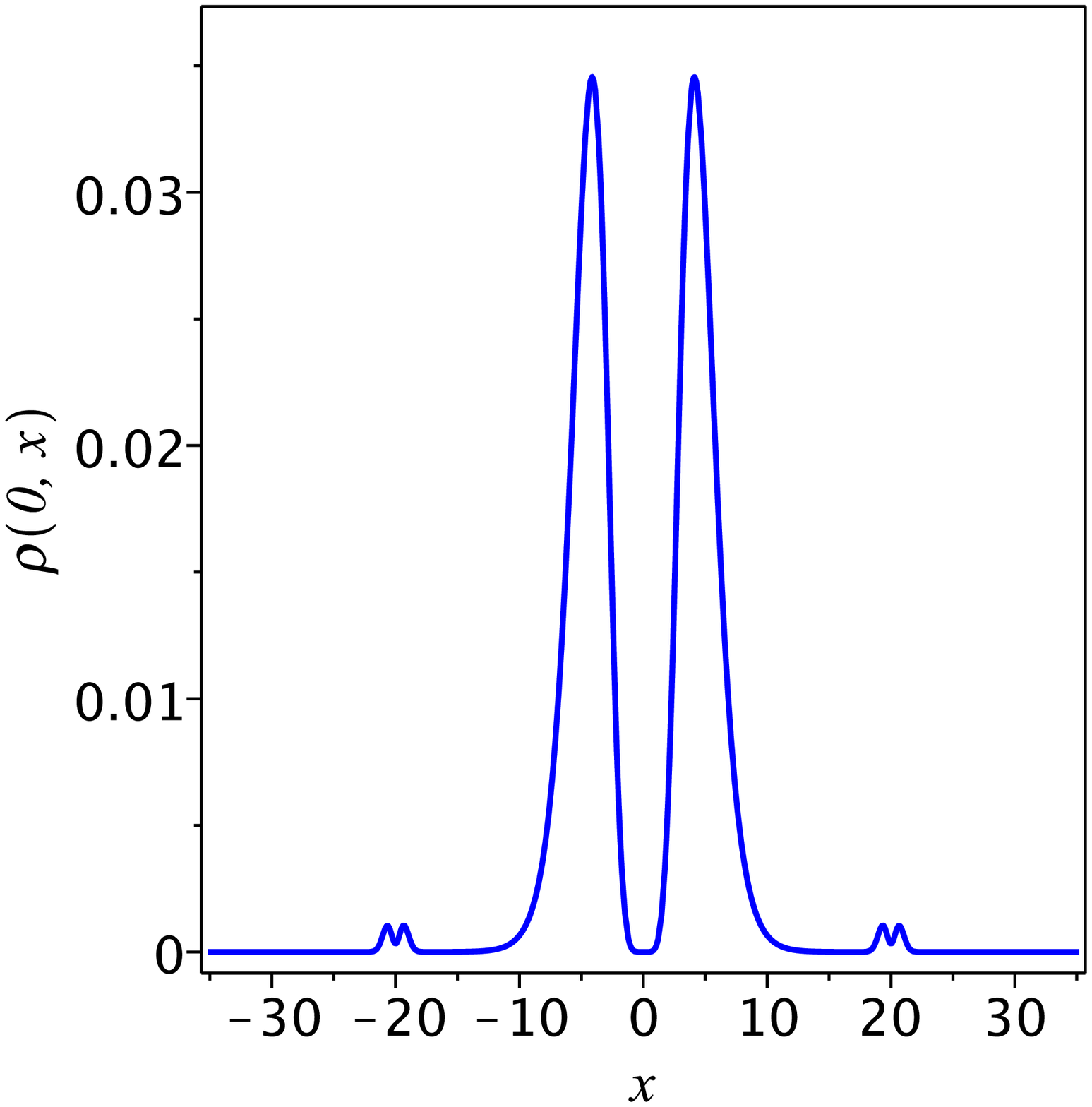}
\includegraphics[scale=0.42]{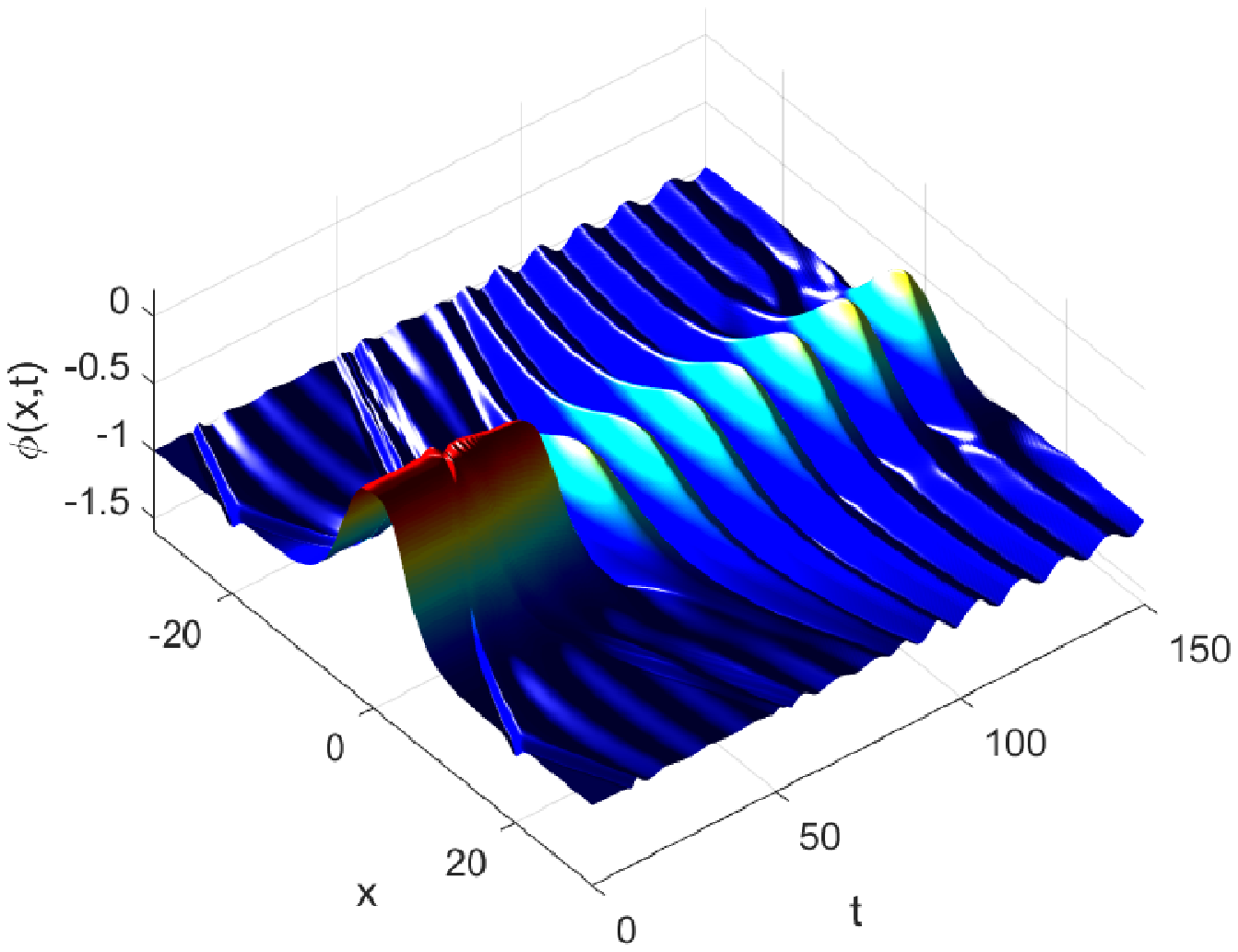}
\includegraphics[scale=0.36]{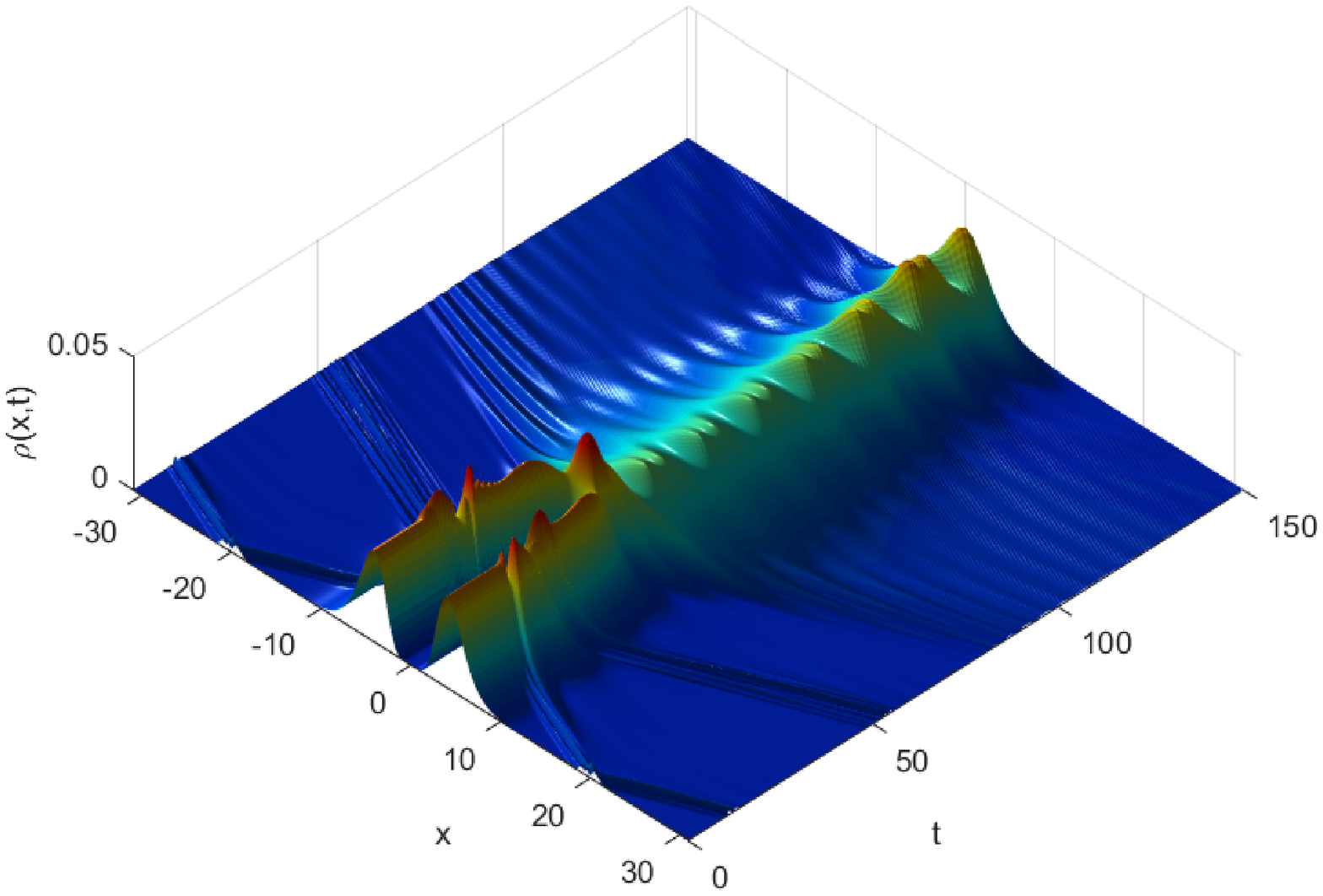}
\caption{Upper panels: initial profile of the perturbed lump-like defect viven by Eq. (\ref{eq4.1}) (left) together with the corresponding energy density (rigth). Here $B_0=0.1$ and the perturbations are locate at $x=\pm 20$). Lower panels: three-dimensional plots of the scalar field and the energy density showin the destruction of the lump defect leaving an oscillon at the origin.}
\end{figure}
\begin{figure}[htb]
\centering
\includegraphics[scale=0.4]{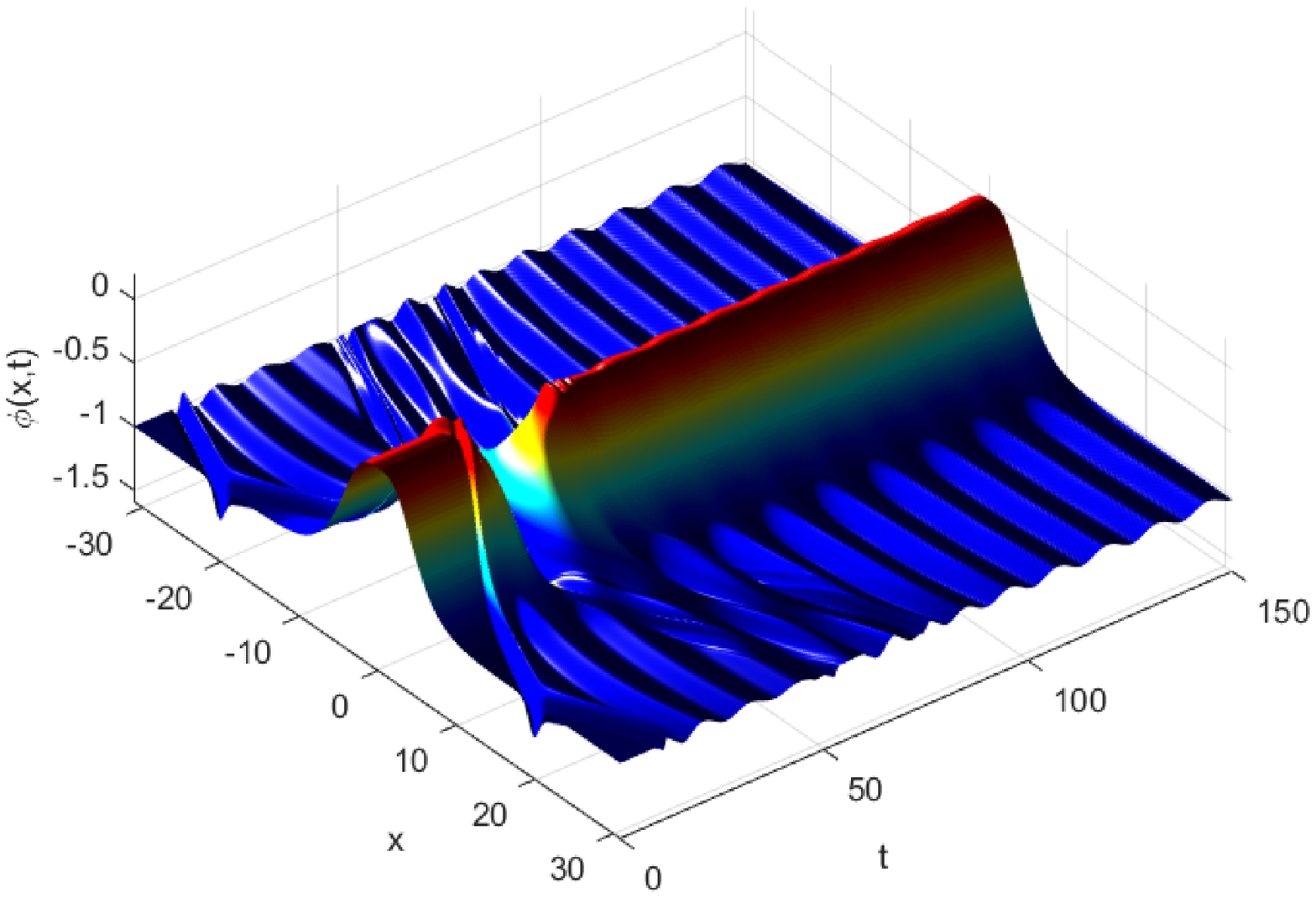}\includegraphics[scale=0.4]{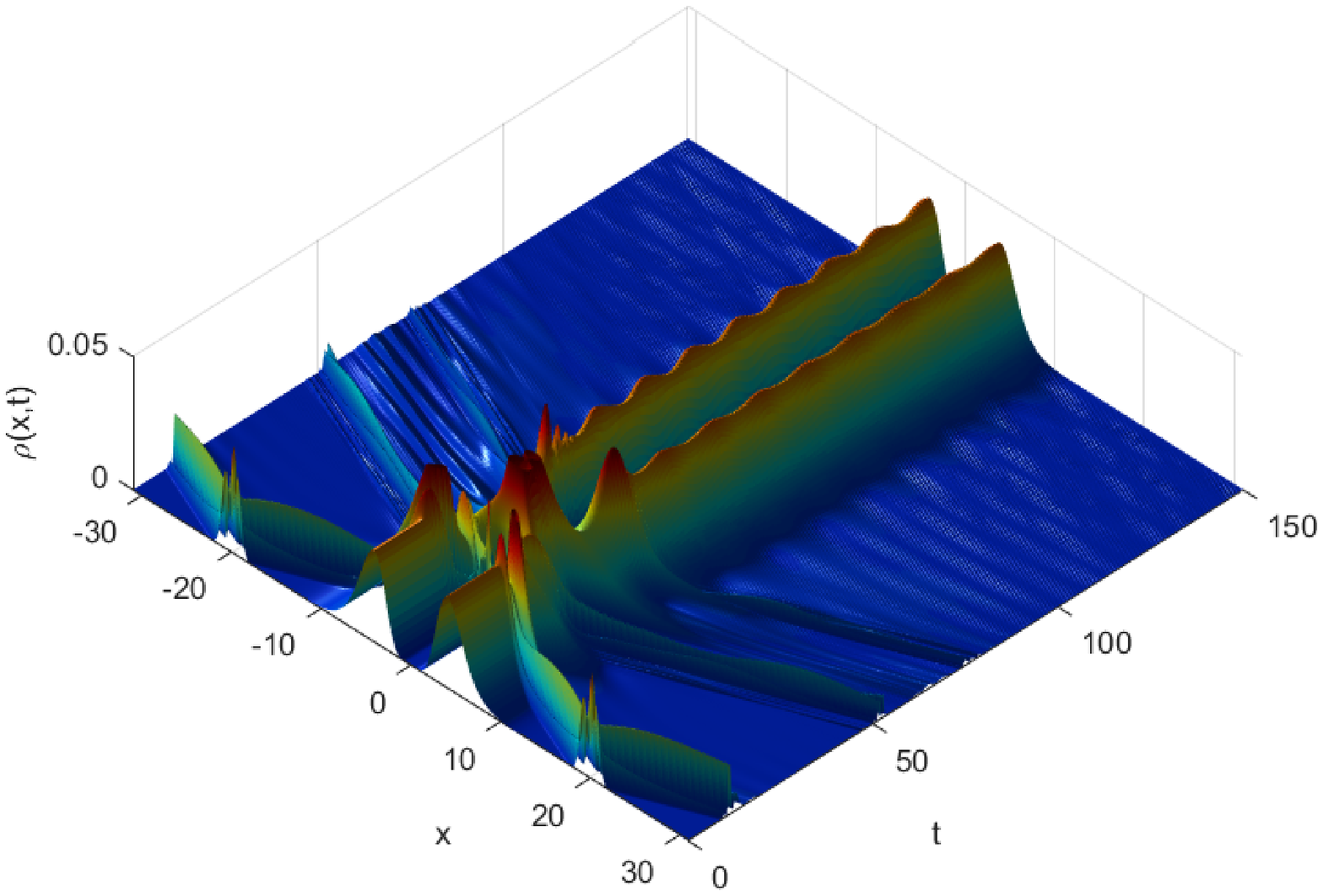}
\includegraphics[width=4.5cm,height=3.5cm]{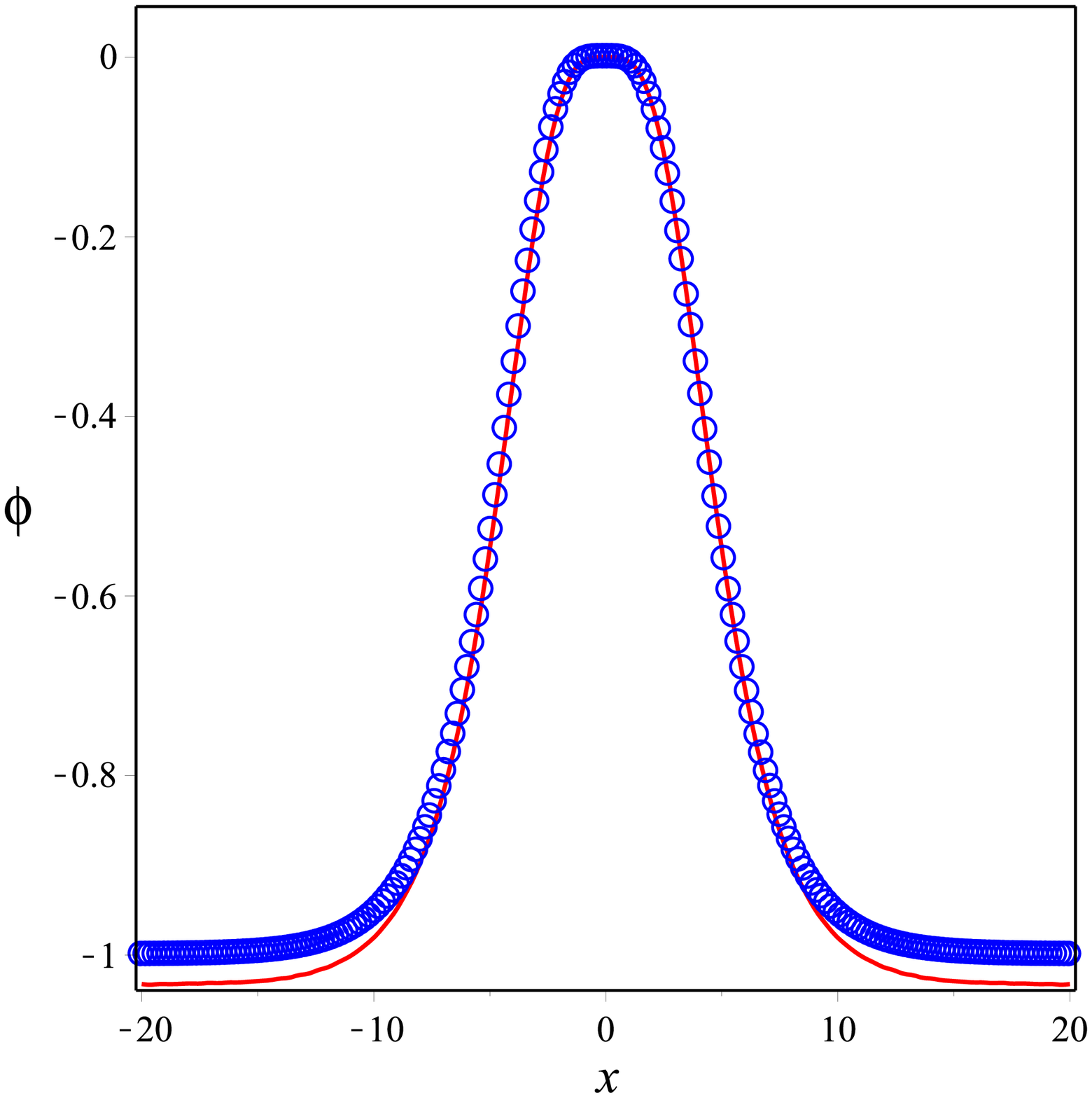}
\hspace{3cm}\includegraphics[width=4.5cm,height=3.5cm]{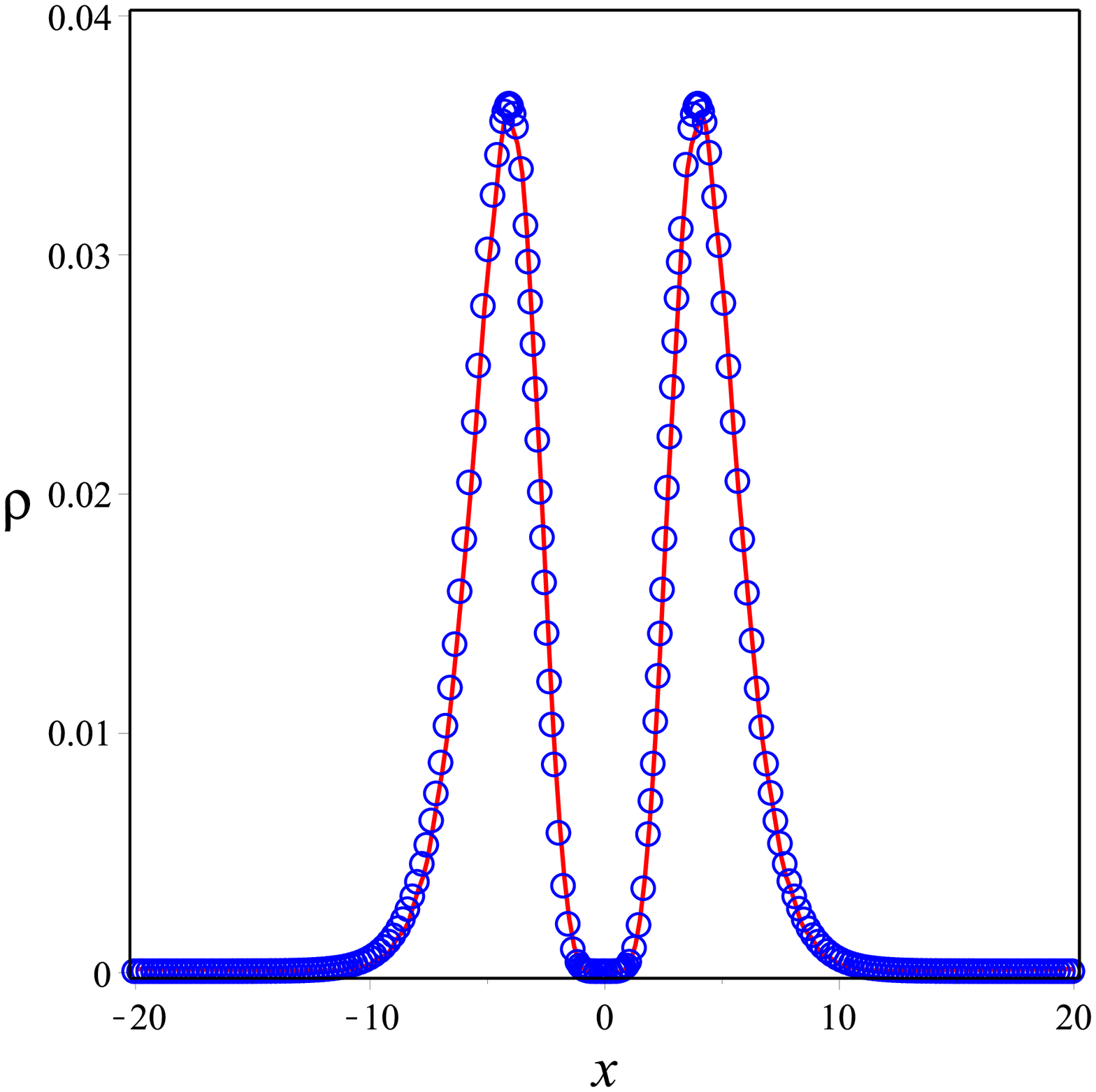}
\caption{Upper panels: critical configuration after perturbing the lump-like defect with $B_0=0.2649705$ viewed from the three-dimensional plots of the scalar field (left) and the energy density (right). Lower panels: the numerical (line) and the exact (circles) profiles of the scalar field and the energy density at $t=150$. The fitting using the expression (3.1) has the following parameters: $A_0 \simeq -1, b_0 \simeq 0.2561$ and $q=4$.}
\end{figure}

\section{Discussion}%

We presented a thorough revision of the collision of a pair of two-kinks as a follow-up of Ref. \cite{tiago} performing more accurate simulations that uncovered new features. In general, the interaction of the two-kinks is strongly inelastic with a considerable amount of scalar field radiated away. The examination of the internal modes of a two-kink reveals that continuum spectrum associated with radiation or boson modes is broader than the corresponding one for a kink as shown in Table 1 that would explain the robust inelastic nature of the collision. The surviving oscillon of the first class of solutions has a small amplitude and eventually a short lifetime if compared with the resulting oscillon of the collision of a pair of kinks \cite{kink_inter_3}. 

We identified the second solution as the formation and escape of a kink-antikink pair moving apart to each other together with an oscillatory component due to the excitation of the vibrational mode. These kinks belong to the quartic interaction potential $U(\varphi)$ (see Eq. (\ref{eq2.4})) with fixed parameters $\kappa_0=-\varphi_0=0.5$. The creation of kinks from the collision of the two-kinks suggests that kinks are fundamental topological defects in the context of all possible two-kinks described by the potential (\ref{eq1.2}). Another example of the creation of kinks takes place when two wave trains representing a sequence of particles collide \cite{train_kinks,train_kinks2}.

The critical configurations are one of the most remarkable features of the interaction of two-kinks. They are generated if the impact velocity assumes very specific or the critical values. We have identified the critical configurations by nontopological lump-like defects ($p$ even) that survives as long as the impact velocity approaches to one of the critical values. Furthermore, the numerical experiments indicated that the critical configurations are the boundaries of the windows of escape similar to the two-bounce windows of the interaction of a pair of kinks. We have found numerical evidence of the fractal structure of the windows of escape that of the Moreover, we have shown the fractal structure of the windows of escape due to the chaotic nature of the collision between the two-kinks.

The last experiment showed that, although the lump-like defects are unstable under small perturbations, they can form a metastable configuration after being perturbated. Therefore, the production of a metastable configuration after selecting a specific value of the parameter representing the amplitude of the perturbation is similar to selecting a critical impact velocity. We may speculate a possible application in Cosmology in which more generic domain walls - two-kinks -  made up of real scalar fields interact producing localized lump-like defects lasting for some time. Some lump-like defects have small lifetimes while others may survive for longer times. Then, on average we have localized energy densities that can play the role of dark matter.

We point out a possible future work involving the dynamics of scalar fields described by the potential (\ref{eq1.2}) in two-dimension for which the case $p=1$ describes the long-lived localized excitations of the scalar field, or merely an oscillon \cite{salmi}. In particular, this case has appeared in several applications in Cosmology \cite{gleiser,amin}. With more generic kinks it is possible potentially new features can emerge.

\acknowledgments
T. S. Mendon\c ca acknowledges the financial support of the Brazilian agency Coordenacao de Aperfei\c coamento de Pessoal de N\'ivel Superior (CAPES). H. P. de Oliveira thanks to Conselho Nacional de Desenvolvimento Cient\'ifico e Tecnol\'ogico (CNPq) and Funda\c c\~ao Carlos Chagas Filho de Amparo \`a Pesquisa do Estado do Rio de Janeiro (FAPERJ) (Grant No. E-26/202.998/518 2016 Bolsas de Bancada de Projetos (BBP)).

\end{document}